\documentclass{article}

\usepackage[english]{babel}

\usepackage[letterpaper,top=2cm,bottom=2cm,left=3cm,right=3cm,marginparwidth=1.75cm]{geometry}

\usepackage{graphicx}%
\usepackage{multirow}%
\usepackage{rotating}
\usepackage{amsmath,amssymb,amsfonts}%
\usepackage{amsthm}%
\usepackage{mathrsfs}%
\usepackage[title]{appendix}%
\usepackage{xcolor}%
\usepackage{textcomp}%
\usepackage{manyfoot}%
\usepackage{booktabs}%
\usepackage{algorithm}%
\usepackage{algorithmicx}%
\usepackage{algpseudocode}%
\usepackage{listings}%
\usepackage{subcaption}
\usepackage{float}
\usepackage{abstract}
\usepackage{url}

\usepackage[colorlinks=true, allcolors=blue]{hyperref}
\usepackage{silence}
\providecommand{\keywordsname}{Keywords} 
\newcommand{\keywords}[1]{\textbf{\keywordsname: }#1}

\title{Lonely individuals show distinct patterns of social media engagement}
\author{Yajing Wang, Talayeh Aledavood, Juhi Kulshrestha}

\begin{document}
\maketitle

\begin{abstract}
Loneliness has reached epidemic proportions globally, posing serious risks to mental and physical health. As social media platforms increasingly mediate social interaction, understanding their relationship with loneliness has become urgent. 
While survey-based research has examined social media use and loneliness, findings remain mixed, and little is known about when and how often people engage with social media, or about whether different types of platforms are differently associated with loneliness.
Web trace data now enable objective examination of these behavioral dimensions. We asked whether objectively measured patterns of social media engagement differ between lonely and non-lonely individuals across devices and platform types.
Analyzing six months of web trace data combined with repeated surveys ($N=589$ mobile users; $N=851$ desktop users), we found that greater social media use was associated with higher loneliness across both devices, with this relationship specific to social media rather than other online activities. 
On desktop, lonely individuals exhibited shorter sessions but more frequent daily engagement.
Lonely individuals spent more time on visual-sharing ($g = -0.47$), messaging ($g = -0.36$), and networking-oriented platforms on mobile. 
These findings demonstrate how longitudinal web trace data can reveal behavioral patterns associated with loneliness, and more broadly illustrate the potential of digital traces for studying other psychological states. 
Beyond research, the results inform the responsible design of digital interventions and platform features that better support psychological well-being across different technological contexts.
\end{abstract}


\section{Introduction}

Loneliness is a major public health issue worldwide~\cite {WHO2025Loneliness}. It is more widespread than common major health issues, such as smoking, diabetes, and obesity, and is becoming a critical concern for social well-being, with comparable levels of risk to health~\cite{WHO2025Loneliness, hhs2023loneliness,holt2021loneliness,ezzatvar2025loneliness,cene2022aha}. Mounting evidence links loneliness to depression, anxiety, cognitive decline, biological stress responses, and even increased risk of premature mortality~\cite{holt2024social,hhs2023loneliness,song2023loneliness,sbarra2023loneliness,liang2024loneliness, zhou2023loneliness}. As social interactions increasingly shift into digital environments, concerns have grown that social media may deteriorate our mental well-being and increase our levels of loneliness~\cite{twenge2017screen,marttila2021social}. Thus, investigating the relationship between online social activity and loneliness has become imperative to address one of the most pressing psychosocial challenges of the digital age.

Loneliness is understood as a subjective emotional experience of social isolation or a perceived lack of meaningful connection with others, reflecting a mismatch between desired and actual social interactions~\cite{Weiss1973Loneliness}. 
According to the World Health Organization, about 16\% of people worldwide experience loneliness~\cite{WHO2025Loneliness}. Similarly, a Gallup World Poll found that nearly 23\% of adults reported feeling lonely for much of the previous day~\cite {Gallup2023}. Among older adults, a recent meta-analysis including over 1.2 million participants estimated that 27.6\% experience loneliness~\cite{Salari2025}. Although levels of loneliness decreased after the COVID-19 pandemic, older adults with fair or poor health have remained higher than pre-pandemic levels even after social restrictions were lifted~\cite {umich2023loneliness}. At the same time, more than 23\% of adolescents and up to 50\% of young adults reported loneliness, indicating that it is no longer confined to older populations\cite{Supke2025Loneliness}.
These findings highlight that loneliness is a global phenomenon that extends beyond temporary crises, cutting across regions, age groups, and cultural contexts. 
Previous research has highlighted several critical factors of loneliness, such as relationship quality and life circumstances~\cite{Weiss1973Loneliness,cacioppo2014evolutionary,luhmann2023loneliness}. However, with social interactions shifting online, especially through social media, scholars have raised concerns about how digital contexts may shape experiences of loneliness~\cite{marciano2024social,WHO2025Loneliness}.

Social media has become an integral part of daily life, with nearly 5.41 billion social media users worldwide as of mid-2025, representing 95.7\% of global internet users~\cite{DataReportal2025SocialMedia}. Prior research has largely examined its associations with mental health conditions such as depression and anxiety, with meta-analytic evidence indicating moderate positive correlations between problematic use and elevated symptoms~\cite{Ahmed2024SocialMediaMeta}. Large-scale studies have also compared social media use among adolescents with and without diagnosed mental health conditions~\cite{Fassi2025MHCYP}. However, loneliness has received far less attention within digital behavior research, despite its recognition as a significant risk factor for both mental and physical health and a pressing public health concern. A long tradition of loneliness research has shown that both the quantity and quality of one’s social relationships are fundamental determinants of loneliness~\cite{Weiss1973Loneliness, Hawkley2010LonelinessMatters, Cacioppo2008Loneliness}. As social interactions increasingly shift into digital environments, with social media often substituting face-to-face interactions and offline connections, investigating its relationship with loneliness has become both timely and necessary.

Most existing studies on social media and mental health rely either on self-reported estimates of use or on questionnaires targeting problematic or addictive use~\cite{marttila2021social,marciano2024social,Turavinina2025MoreOnline,macdonald2022loneliness}.
However, such self-reports can be highly inaccurate, with fewer than 10\% of estimates falling within 5\% of actual logged behavior, and some individuals overestimating their screen time by more than 60\%~\cite{Andrews2015BeyondSelfReport}. 
Moreover, while instruments such as the Bergen Social Media Addiction Scale~\cite{Andreassen2012BFAS} and the Social Media Disorder Scale~\cite{vandenEijnden2016SMDS} have demonstrated good psychometric properties, they target problematic use. Therefore, they cannot be readily applied to study the relationship between everyday social media use and loneliness.

Some studies have attempted to overcome these limitations by incorporating objective measures of digital activity to study the impact of online activity on individuals' well-being.
Yet these studies typically focus on a single platform, such as Facebook or Instagram \cite{Verduyn2015PassiveFacebook, Sherlock2019InstagramWomen, Zhao2023TelematicsInstagram}. 
This narrow scope overlooks the fact that individuals usually maintain multiple social media accounts and routinely switch between platforms and devices~\cite{Reeves2019Screenomics}. 
For instance, people may use WhatsApp to communicate with close friends, Instagram to share personal updates, and YouTube for entertainment. 
Ignoring this cross-platform context risks misrepresenting the diversity of social media use and its potential links to well-being.

Our study offers evidence-based insights into the relationship between everyday social media use and feelings of loneliness. We analysed data from a six-month panel study that combined repeated monthly survey measures of loneliness (UCLA Loneliness Scale, 3-item version~\cite{Russell1996UCLA,Hughes2004ShortUCLA}) with passively collected web trace data spanning both mobile and desktop devices. 
People often use these devices in different ways~\cite{Reeves2019Screenomics,Dear2008}. Mobile devices are often used for messaging and social networking~\cite{Heitmayer2022VideoDiary,Jokela2015}. In contrast, desktops are more commonly associated with information-seeking or work-related tasks~\cite{Heitmayer2022VideoDiary, Jokela2015, Dear2008}. Therefore, we analyse mobile and desktop traces separately. This design links validated psychological scales with objective, cross-platform measures of web use, while accounting for the differences in online engagement between devices.

Previous research has examined the links between social media and well-being~\cite{marttila2021social,marciano2024social,Turavinina2025MoreOnline,macdonald2022loneliness,Verduyn2015PassiveFacebook, Sherlock2019InstagramWomen, Zhao2023TelematicsInstagram}, but the evidence regarding loneliness remains mixed~\cite{primack2017social,hunt2018FOMO,roberts2024,nowland2018,wang2024,matthews2025}. Some studies report that greater time spent on social media is associated with higher loneliness~\cite{primack2017social,hunt2018FOMO,roberts2024}, while others find no relationship or even potential benefits from online connection~\cite{nowland2018,wang2024,matthews2025}. 
Based on this literature, we formulated our first hypothesis (H1): Social media use is associated with loneliness.

Beyond the overall amount of social media use, much less is known about its temporal patterns, such as how frequently people engage with these platforms and whether such rhythms matter. Initial evidence from digital behavior research suggests that rhythms of people's online activities can reflect psychosocial functioning~\cite{torous2021growing, aledavood2025multimodal}, but their relevance for loneliness has not been systematically assessed. Thus, we proposed our second hypothesis (H2): Temporal patterns of social media engagement differ between lonely and non-lonely individuals. 

Finally, prior work suggests heterogeneity across platforms, with different services and affordances potentially influencing well-being in distinct ways~\cite{valkenburg2022}. However, systematic comparisons across platform types remain limited. 
Therefore, we advanced our third hypothesis (H3): Patterns of specific types of social media use differ between lonely and non-lonely individuals.

\section{Results}\label{sec2}

Our study design combines web trace data and repeated surveys containing validated psychological scales for measuring loneliness (UCLA-3~\cite{Hughes2004ShortUCLA}), collected from a panel of participants (N = 589 mobile users, N = 851 desktop users) over a six-month period. Participants completed an online questionnaire once per month, resulting in six measurement waves that we aligned with their web trace data (see Figure~\ref{fig:design} for an overview of the data collection process).  
Web trace data included mobile and desktop use logs, which recorded the app or domain visited, the time of the visit, and the duration of the visit. From these web traces, we extracted the visits to social media platforms and computed features that captured different aspects of social media use, such as the total time spent on social media, temporal patterns of social media engagement, and categories of social media visited. We analyzed the mobile and desktop datasets separately, as existing research shows that online behavior differs across devices~\cite{Heitmayer2022VideoDiary, Jokela2015, Dear2008}.

\subsection{Higher social media use associated with greater loneliness}

To test our primary hypothesis (H1) that social media use is associated with loneliness, we examined the relationship between the average time spent on social media per day and the loneliness scores using linear mixed-effects models (LMMs). 

In the first model, we included time spent on social media, total time spent online and individuals' sociodemographics as covariates (see model M1 in Section~\ref{sec:statanalysis} for detailed model specifications). 
We found that the daily average time spent on social media was positively associated with loneliness across both mobile and desktop devices (mobile: $b = 0.085$, $P = 0.034$; desktop: $b = 0.102$, $P = 0.016$). This positive association means that participants who spent more time on social media per day reported higher loneliness. 
This relationship was significant after controlling for total time spent online and sociodemographic covariates (age, gender, income, chronotype,
and relationship status; for more model details, see Section~\ref{sec:statanalysis}). Loneliness scores were lower among partnered individuals, individuals with higher income ($P = 0.055$, with only marginal effects in mobile), and morning-type individuals across both devices.
Loneliness scores of mobile users showed a slight decrease across the wave, but no such association was observed for desktop users. Desktop users showed a negative association between loneliness and age, while gender showed no significant effect across devices. 
The model(M1) showed substantial between-person variability (mobile: $\tau_0 = 1.75$; desktop: $\tau_0 = 1.64$), with intra-class correlations of 0.68–0.70. While fixed effects explained a modest proportion of variance (marginal $R^2 = 0.09$ for both datasets), the full models including random effects explained considerably more variance (conditional $R^2 = 0.72$ for mobile and $0.71$ for desktop; see full model output in Appendix \ref{secA1}).

Next, to determine whether this association with loneliness is specific to social media use or reflects broader web use, we compared social media use against other online activities (entertainment, games, health-related, news, search, shopping, and productivity). We extended the previous LMM (Model M1) to also include the time spent on each of these online activities as independent variables in addition to time spent on social media (see model M2 in Section~\ref{sec:statanalysis} for detailed model specifications). 
Social media remained the only online activity significantly associated with loneliness, even when we tested against a range of online activities. Desktop users showed a significant positive association between social media use and UCLA-3 loneliness scores ($b = 0.004, P = 0.013$), while mobile users exhibited a weaker, marginally significant effect ($b = 0.003, P = 0.082$). No other online activity was found to be significantly associated with loneliness across either device. Sociodemographic covariates maintained consistent effects as observed in model M1. Similarly, the extended model M2 demonstrated between-person variability (random intercept variance $\tau_0 = 1.76$ for mobile, $\tau_0 = 1.65$ for desktop). Fixed effects explained a modest proportion of variance (marginal $R^2 = 0.09$ for both). In contrast, full models including random effects accounted for considerably more variance (conditional $R^2 = 0.72$ for mobile, 0.71 for desktop; see full model output in Appendix~\ref{secA1}).

Lastly, to explore whether the relationship between social media and loneliness varies across different population groups, we conducted subgroup analyses by running model M1 separately for participants in each subgroup of the demographic characteristics (age, gender, income, chronotype, and relationship status). 
The association between social media use and loneliness varied across sociodemographic subgroups. Social media use was significantly associated with higher loneliness among morning type mobile users ($P = 0.021$), middle-aged mobile users (31--60 years, $P = 0.006$),  senior desktop users (60+, $P = 0.011$), high-income desktop users ($P = 0.039$), intermediate chronotype desktop users ($P = 0.001$), and both female mobile ($P = 0.021$) and female desktop ($P = 0.038$) users. 
No significant associations were observed for the younger age group (18--30), evening type, or male users across either devices. 
These findings indicate that the relationship between social media use and loneliness is not uniform across the population
 (see detailed information and model results in Appendix~\ref{tab:subgroup-summary-socialmedia}).

\subsection{Temporal patterns of social media engagement differ between lonely and non-lonely individuals}

After establishing that more social media use is associated with higher loneliness (H1), we next tested our second hypothesis that temporal patterns of social media engagement differ between lonely and non-lonely individuals (H2). Therefore, we computed four temporal social media use metrics (session duration, daily sessions, inter-session intervals, and difference in time spent on social media during work and free hours; definitions in Section~\ref{sec:quantSM}). We compared these temporal metrics between lonely participants ($score > 5$, $n = 160$ for mobile, $n = 225$ for desktop) and non-lonely participants ($score = 3$, $n = 131$ for mobile, $n = 213$ for desktop) using equivalence testing alongside traditional hypothesis testing, using 0.4 as the smallest effect size of interest (SESOI).

\begin{figure}[htbp]
    \centering
    \includegraphics[width=1\linewidth]{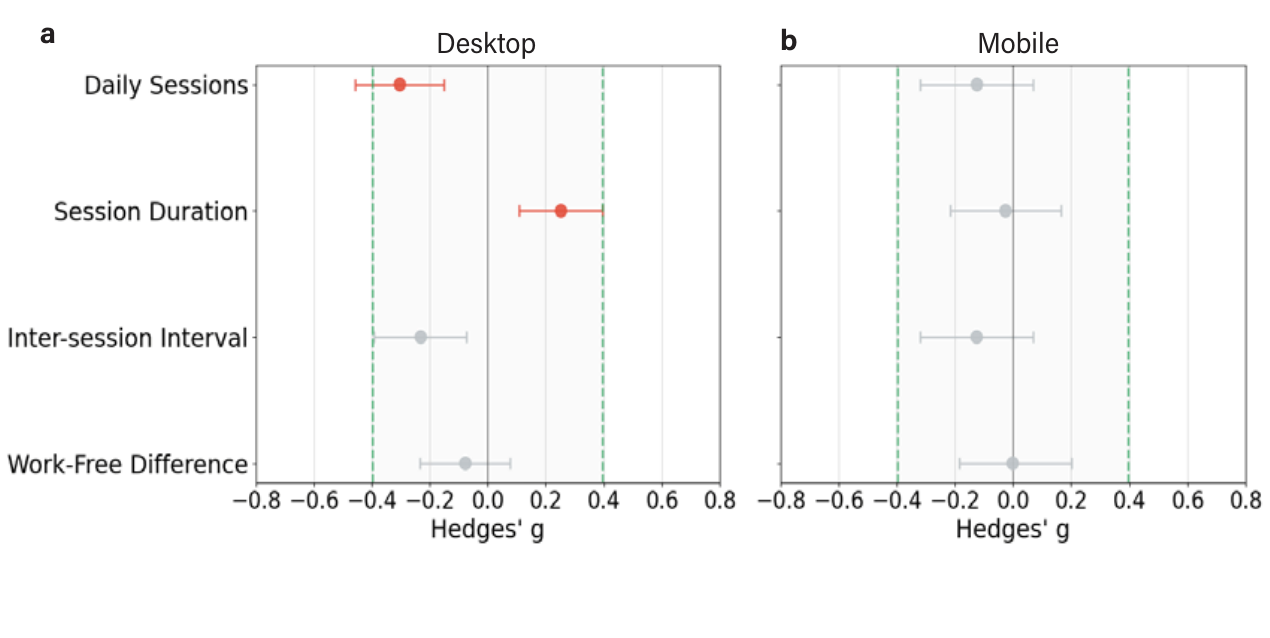}
   \caption{\textbf{Effect sizes for desktop and mobile temporal patterns comparing lonely and non-lonely individuals.} Effect sizes are presented as Hedges' g with 90\% confidence intervals (error bars). Negative values indicate higher values in the lonely group; positive values indicate higher values in the non-lonely group. Red points indicate effects exceeding the SESOI ($\mid g\mid\geq0.4$). \textbf{a}, Desktop use showed meaningful differences in daily sessions and session duration. \textbf{b}, Mobile use showed no meaningful differences across all metrics.}

    \label{fig:H2}

\end{figure}

The Figure~\ref{fig:H2}\textbf{a} shows effect sizes (Hedges' g) for comparison of lonely and non-lonely individuals on desktops. 
Negative values indicate lonely individuals have a higher value for the feature, while positive values indicate non-lonely individuals demonstrate higher values.
Overall, we found that lonely individuals exhibited a fragmented social media use pattern spread across the day. 
They initiated more number of sessions each day (3.72 vs.\ 3.11; $g = -0.30$, 90\% CI [$-0.46$, $-0.15$], $P = 0.006$), yet each session duration was shorter (4.23 vs.\ 6.93 hours; $g = 0.25$, 90\% CI [$0.11$, $0.40$], $P = 0.018$), with correspondingly shorter intervals between sessions showing a smaller but significant effect (0.58 vs.\ 0.71 hours; $g = -0.23$, 90\% CI [$-0.39$, $-0.07$], $P = 0.023$). 
The difference in social media time during work and free hours did not differ significantly between lonely and non-lonely individuals ($P = 0.424$). 

The Figure~\ref{fig:H2}\textbf{b} shows the effect sizes (Hedges' g) for comparison of lonely and non-lonely individuals on mobile devices. 
None of the temporal social media use features differed significantly between groups. Full descriptive statistics and test results are reported in Supplementary Table~\ref{tab:H2_Full}. 

\subsection{Types of social media use differ between lonely and non-lonely individuals}

\begin{figure}[htbp]
    \centering
    \includegraphics[width=1\linewidth]{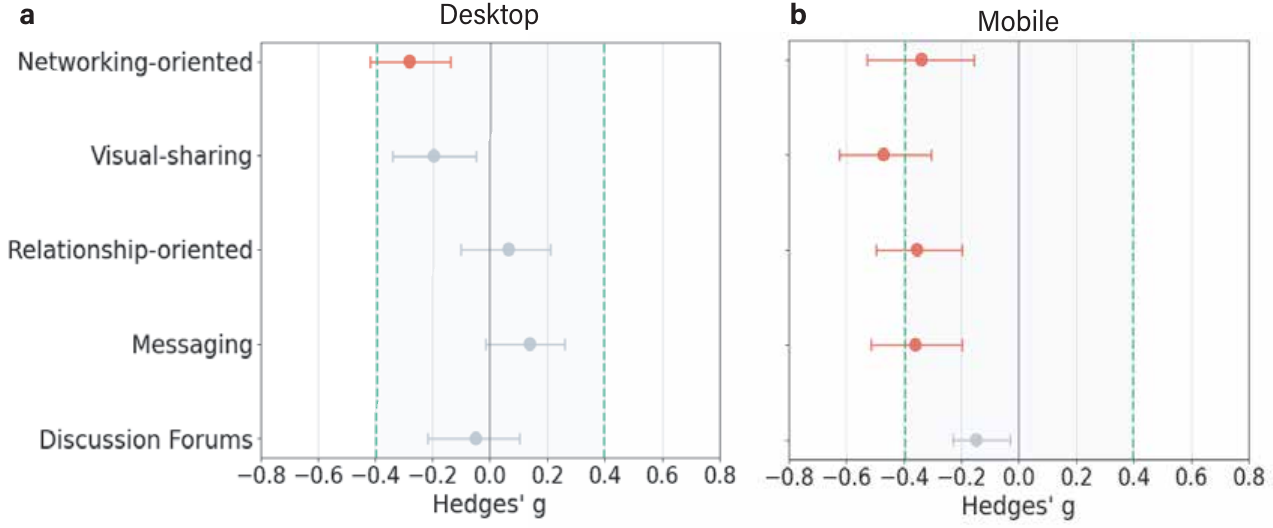}
   
    \caption{\textbf{Effect sizes for types of social media use comparing lonely and non-lonely individuals.} Effect sizes are presented as Hedges' g with 90\% confidence intervals (error bars). Negative values indicate higher use in the lonely group; positive values indicate higher use in the non-lonely group. Red points indicate effects exceeding the SESOI ($\mid g\mid\geq0.4$). \textbf{a}, Desktop use showed a meaningful difference in networking-oriented platforms. \textbf{b}, Mobile use showed meaningful differences across networking-oriented, visual-sharing, relationship-oriented, and messaging platforms.}
    \label{fig:H3}
\end{figure}

To test our third hypothesis that patterns of specific types of social media use differ between lonely and non-lonely individuals(H3). We compared average time spent daily on five distinct types of social media platforms (visual-sharing, messaging, networking-oriented, relationship-oriented, and discussion forums) between lonely and non-lonely individuals using the same group definitions as in our temporal analysis. Figure~\ref{fig:H3} presents group comparison results in time spent on different types of social media. Negative effect sizes indicate greater use by lonely individuals.

On desktop (Figure~\ref{fig:H3}\textbf{a}), lonely individuals spent significantly more time on networking-oriented platforms(12.1 vs.\ 6.7 minutes; $g = -0.28$, 90\% CI [$-0.42$, $-0.14$], $P = 0.015$). Visual-sharing platforms showed a small difference between groups (13.0 vs.\ 9.6 minutes; $g = -0.20$, 90\% CI [$-0.34$, $-0.05$], $P = 0.094$). No significant differences were observed for the remaining variables.

On mobile(Figure~\ref{fig:H3}\textbf{b}), lonely individuals spent significantly more time on networking-oriented platforms(18.4 vs.\ 10.9 minutes; $g = -0.34$, 90\% CI [$-0.53$, $-0.16$], $P = 0.004$), visual-sharing platforms(34.2 vs.\ 16.6 minutes; $g = -0.47$, 90\% CI [$-0.62$, $-0.30$], $P < 0.001$), relationship-oriented platforms(2.1 vs.\ 0.9 minutes; $g = -0.35$, 90\% CI [$-0.50$, $-0.20$] $P = 0.003$), messaging platforms(22.4 vs.\ 15.8 minutes; $g = -0.36$, 90\% CI [$-0.51$, $-0.20$]; $t = -3.24$, $P = 0.003$). Discussion forums showed no significant difference($P = 0.168$).The corresponding descriptive statistics and test results are provided in the Supplementary Table~\ref{tab:H3_Full}.

\subsection{Sensitivity Analyses Results}

Our findings are robust across sensitivity analyses, though some minor variations were observed. 
Sensitivity analyses using alternative thresholds (20\% and 30\% for survey-taking behavior; 15 and 25 days for mobile activity) yielded consistent results for the main social media-loneliness association. 
Z-score outlier removal emerged as the most stable preprocessing approach (average deviation: 0.048), outperforming other methods, including winsorization and transformations. Statistical test agreement between parametric and non-parametric approaches was high, with perfect concordance (100\%) for mobile device analyses and 57-100\% agreement for desktop analyses. Significant effects remained consistent across preprocessing methods, with variations occurring only for borderline cases (original p-values $0.03-0.06$). The pattern of practically meaningful differences (exceeding SESOI bounds) remained unchanged across all sensitivity analyses. (Preprocessing method stability results are provided in Appendix~\ref{secA3}.)

Our six-month longitudinal analysis of web trace data and self-reported loneliness measures showed three key findings about the relationship between social media use and loneliness. First, social media time was consistently associated with higher loneliness scores across both mobile and desktop platforms, with this relationship remaining significant even when controlling for total screen time and socio-demographic factors. 
When tested against multiple online activity categories, social media emerged as the only digital behavior significantly linked to loneliness, and this association varied across demographic subgroups, particularly affecting morning types, middle-aged adults, seniors, high-income users, and female participants. 
Second, temporal engagement patterns showed device-specific associations with loneliness. On desktop devices, lonely individuals had significantly shorter session durations but more frequent daily sessions compared to non-lonely users. 
Third, lonely participants exhibited significantly greater engagement with social networking platforms across both devices, particularly visual-sharing, messaging, and networking-oriented platforms. 
Together, these findings demonstrate that the relationship between social media use and loneliness is complex and multifaceted, varying by device type, platform category, and individual characteristics.

\section{Discussion}\label{sec3}

In this longitudinal study, we examined the relationship between objective social media use and loneliness using digital trace data from both mobile and desktop devices over a six-month period. Our findings described a comprehensive picture of how digital social behavior relates to psychological well-being. We found that social media use was consistently associated with higher loneliness scores across platforms, yet this relationship exhibited device-specific variations in temporal engagement patterns. Lonely individuals demonstrated distinct usage behaviors, characterized by more frequent but shorter desktop sessions, greater engagement with networking-oriented platforms across both devices and visual-sharing, relationship-oriented, and messaging apps on mobile devices. These results challenge simplistic assumptions about the impact of screen time on loneliness and highlight the importance of considering technological context, usage patterns, and platform characteristics when examining digital social behavior and mental health outcomes. 

Our findings report a consistent positive association between social media use and loneliness across both mobile and desktop devices, supporting our first hypothesis (H1). This relationship emerged after controlling for total screen time, indicating that social media engagement, rather than general digital device use, drives the effect. 
The social displacement hypothesis~\cite{nie2002impact, kraut1998internet, Turavinina2025MoreOnline} provides a plausible explanation, suggesting that online engagement may reduce face-to-face interactions and shared activities that protect against loneliness. 
Previous research examining broad screen time has produced mixed findings~\cite{burns2025pandemic,nagata2024screen,qirtas2023relationship,macdonald2022loneliness,twenge2021not}, with some studies reporting associations with psychological well-being~\cite{burns2025pandemic,nagata2024screen,qirtas2023relationship}, and others finding weak or inconsistent effects~\cite{macdonald2022loneliness,twenge2021not}. 
By isolating social media, our results help explain why previous studies yielded mixed findings and point to the value of activity-specific approaches to digital well-being. 
Subgroup analyses indicated boundary conditions.
The association was strongest among morning types and middle-aged adults, consistent with evidence that chronotypes shape social network structures~\cite{aledavood2018chronotypes}.
By contrast, effects were weaker or absent in groups where loneliness is already strongly shaped by structural vulnerabilities, such as being single, having lower income, or belonging to younger age groups~\cite{carr2010advances,proulx2007marital,lu2022moderating,davis2025associations,lok2025chronotype,chang2024associations}. These patterns highlighting the multifactorial nature of loneliness and the importance of considering individual and contextual factors in understanding digital media effects.

Our analysis of temporal engagement patterns revealed device-specific differences in how loneliness relates to social media usage behaviors, providing mixed support for our second hypothesis (H2). 
The absence of significant temporal differences between lonely and non-lonely individuals on mobile devices may reflect several factors related to the nature of mobile technology use. Mobile devices have become deeply integrated into daily routines, serving as essential tools for constant connectivity, communication, and micro-interactions throughout the day~\cite{fortunati2022smartphone,vorderer2017permanently}. 
This ubiquitous and flexible integration, spanning diverse locations, situations, and social contexts~\cite{schrock2015communicative,fortunati2022smartphone,vorderer2017permanently}, may create relatively uniform usage patterns across individuals and introduce variability that obscures systematic differences between loneliness groups. Mobile social media interactions, which are often brief and fragmented, may operate through different psychological processes than the extended and intentional use typical of desktop devices~\cite{Reeves2019Screenomics}. 
In contrast, desktop usage patterns revealed clear distinctions between lonely and non-lonely participants, with lonely individuals engaging in shorter but more frequent social media sessions. 
These shorter sessions may reflect superficial or unsatisfying interactions that lack the sustained engagement needed to build deeper social connections. 
The higher frequency of sessions suggests repeated attempts to seek social contact, reflecting a cycle of seeking but not finding the supportive relationships that effectively buffer against loneliness.
Desktop environments, typically associated with more intentional and focused activities rather than everyday social media use, allow temporal patterns of social media engagement to better capture group differences compared to the more habitual social media use of mobile devices.

Our results about social media usage types showed that lonely individuals demonstrated significantly greater engagement with visual-sharing, messaging, networking-oriented, and relationship-oriented platforms, supporting our third hypothesis (H3). 
These differences were more significant on mobile devices compared to desktop devices, where only networking-oriented platforms showed clear distinctions between loneliness groups.
The social comparison framework provides a convincing explanation: visual-sharing, networking, and friendship-oriented platforms expose users to curated self-presentations, amplifying upward comparisons that reinforce feelings of inadequacy and isolation~\cite{Leon1954comparison, Vogel2014SocialComparison, Verduyn2020SNSComparison}. 
While such platforms are useful for maintaining broad networks and weak ties, they typically provide less emotional support than close relationships~\cite{Wellman1990DifferentStrokes, Rinderknecht2023Tie}. 
Heavy engagement may thus reflect attempts to compensate for insufficient strong ties, in line with prior findings that these platforms can displace, rather than complement, close social connections~\cite {primack2017social}. 
The stronger type-specific effects on mobile devices likely reflect their integration into daily routines. 
Mobile devices facilitate more frequent, spontaneous interactions throughout the day, making use choice more revealing of underlying social needs, and amplifying both compensatory use and social comparison.
By contrast, type-specific effects were weaker on desktop, reflecting the more intentional and less frequent nature of its use.
Our group comparisons (H2 \& H3) results imply that the habitual integration of mobile devices makes social media type choice more reflective of user motivations and psychological states, whereas the more intentional character of desktop use renders temporal patterns a stronger marker.

Our study contributes to digital well-being research in several ways. 
Methodologically, we combine objective, cross-platform web trace data with repeated survey measures of loneliness, providing a more accurate and comprehensive view of real-world social media use than self-report measures alone. This approach enables us to examine both mobile and desktop behaviors in naturalistic settings while capturing the social media-loneliness relationship over time. 
Empirically, our findings provide several important insights about the relationship between social media use and loneliness. First, we report that social media engagement, rather than general screen time, is associated with loneliness, highlighting the importance of activity-specific measures of technology use over broad measures. 
Second, we observe device-specific differences in how loneliness relates to digital behavior. 
Temporal engagement patterns (such as daily sessions and session duration) distinguish lonely from non-lonely participants on desktop devices. At the same time, platform category preferences emerge as more important factors on mobile devices, with visual-sharing, messaging, networking-oriented, and relationship-oriented platforms showing stronger associations with loneliness in mobile environments.
Theoretically, these findings support both social displacement and social comparison frameworks with demographic and contextual boundary conditions. Our results demonstrate that the relationship between social media use and loneliness is complex, involving multiple interacting factors that vary by device type, platform category, usage patterns, and individual characteristics. 
This complexity highlights the need for comprehensive theoretical models that account for technological affordances, usage contexts, and individual differences in understanding how digital social behaviors impact psychological well-being.

These findings offer several insights for researchers, individuals, and policymakers concerned with digital well-being. 
For researchers, our study highlights the value of combining web trace data with repeated surveys to capture naturalistic behaviors while overcoming limitations of self-report bias and cross-sectional designs. 
This integrative approach provides valuable evidence for associations between specific digital behaviors (social media use) and psychological outcomes (loneliness), and can be extended to examine causal mechanisms in future experimental or intervention studies.
For individuals, the results emphasize that higher social media use is consistently associated with greater loneliness, particularly on visual-sharing and networking platforms. While these associations do not establish causation, they suggest that individuals might benefit from reflecting on their social media engagement patterns and considering how different platforms and usage behaviors relate to their psychological well-being and time allocation for offline relationships.
For social media platform designers, these findings suggest correlations between specific platform characteristics and experiences of loneliness. Understanding these patterns can inform design decisions that promote more meaningful social connections, such as features that encourage deeper engagement rather than passive consumption or superficial interactions.
For policymakers, the results suggest that the amount, patterns, and type of social media use are all associated with loneliness outcomes. Public health initiatives and digital literacy programs can incorporate these findings to help individuals develop healthier relationships with social media. Meanwhile, regulatory considerations might incorporate these platform-specific associations when developing digital wellness guidelines. 
Together, these implications suggest that our evidence-based approaches, informed by objective behavioral data, can guide efforts to optimize digital technologies for promoting psychological well-being.

Our data came from a panel of gig workers who frequently participate in online surveys. To minimize potential biases associated with professional survey participation, we implemented exclusion criteria for participants showing disproportionately high survey-taking activity. Prior research has demonstrated that browsing data from such panels reflects the most visited websites in Germany and that participants' privacy attitudes are comparable to those of non-tracked populations~\cite{Kulshrestha2021WebRoutineness, Stier2020PopulistAttitudes}, supporting the representativeness of web trace data from these samples. 
Due to the practical challenges inherent in collecting large-scale digital trace data, our study focused on participants in Germany. This geographic focus, while enabling deep longitudinal tracking, limits generalizability across different cultural contexts where social media use patterns and experiences of loneliness may vary. Future research extending this approach to diverse cultural settings will be valuable for testing the robustness of these associations. 
For loneliness, we employed the short UCLA-3 scale rather than the full 20-item version. While this shorter measure does not capture the complete breadth of loneliness experiences, the UCLA-3 has been extensively validated as an efficient and reliable screening tool in large-scale studies~\cite{Hughes2004ShortUCLA, Russell1996UCLA}. This choice also helped reduce participant burden and minimize the risk of response fatigue across our six-month repeated measurement design~\cite{Hughes2004ShortUCLA, Burisch1997TestLength}. 
Finally, our digital trace data provided precise measures of usage duration and timing. Still, they did not include information about specific content viewed or interaction types (such as passive consumption versus active engagement). While this approach provides objective behavioral measurements, future research integrating content-level or qualitative measures could offer deeper insights into the mechanisms linking different forms of social media engagement with loneliness.

Using objective web trace data over a six-month period, our study reveals that social media use, rather than general screen time, is associated with loneliness in ways that vary by device, platform characteristics, and individual factors. 
The integration of behavioral data with longitudinal survey measures provides empirical foundations for interventions, platform design innovations, and policy considerations that aim to optimize digital technologies for psychological well-being.

\section{Methods}\label{sec4}
\subsection{Data collection and participant selection}
\begin{figure}[htbp]
    \centering
    \includegraphics[width=\linewidth]{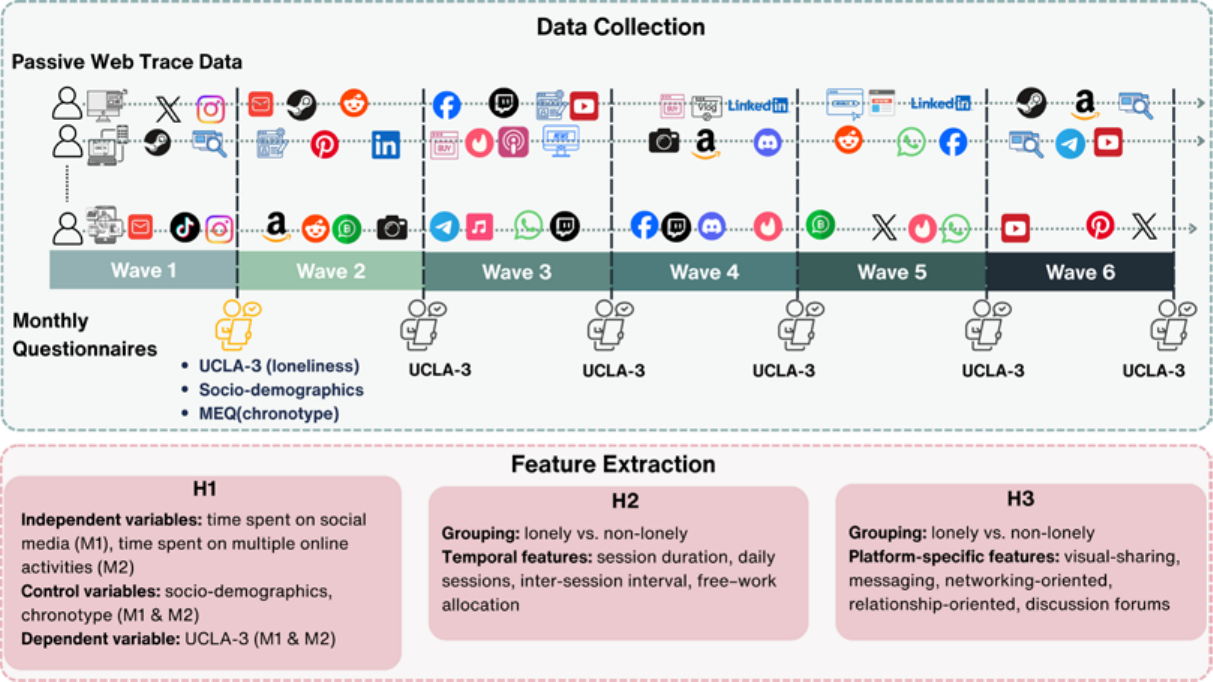}
    \caption{Mixed-method longitudinal study design for investigating digital behavior and loneliness. 
    The study employed a 6-month longitudinal panel design combining continuous objective data collection with loneliness assessment. \textbf{Upper panel}: objective web trace data were continuously collected via desktop web browsing logs and mobile application usage logs, and participants completed monthly online surveys across six waves (Wave 1: UCLA-3 + demographics + MEQ; Waves 2--6: UCLA-3 only). \textbf{Lower panel}: extracted features for each hypothesis. $N = 1,490$, Germany, 2023--2024.}
    \label{fig:design}
\end{figure}

We conducted a longitudinal panel study in Germany between 2023 and 2024, combining passively collected fine-grained web trace data with repeated monthly online surveys over a six-month period. 
Web trace data included URL-level browsing logs for desktop users and application-level usage logs for mobile users, collected continuously throughout the study period using a browser extension on desktop and an application on mobile. 
Each survey wave was associated with a four-week measurement window preceding the survey completion date.
Participants completed UCLA-3\cite{Hughes2004ShortUCLA} once per month, resulting in six waves of loneliness measures. 
In the first wave, participants also provided socio-demographic information, including age, gender, income level, relationship status, and chronotype preference (Figure~\ref{fig:design}).

Participants were recruited through a GDPR-compliant online panel provider \textit{Bilendi GmbH}. Of the 1,490 individuals who completed the initial questionnaire, we excluded one participant who reported a non-binary gender and 52 participants who reported an ``other'' income category, resulting in 1,437 baseline respondents. Attrition was moderate, with 1,107 participants remaining by wave 6 (23\% dropout). The sample was balanced by gender and included diverse age and income groups. Average loneliness scores were relatively stable, decreasing slightly from 4.47 ($SD = 1.65$) in wave 1 to 4.20 ($SD = 1.60$) in wave 6 (see Appendix~\ref{secA4} for complete sample characteristics by dataset).

\begin{figure}[htbp]
    \centering
    \includegraphics[width=\linewidth]{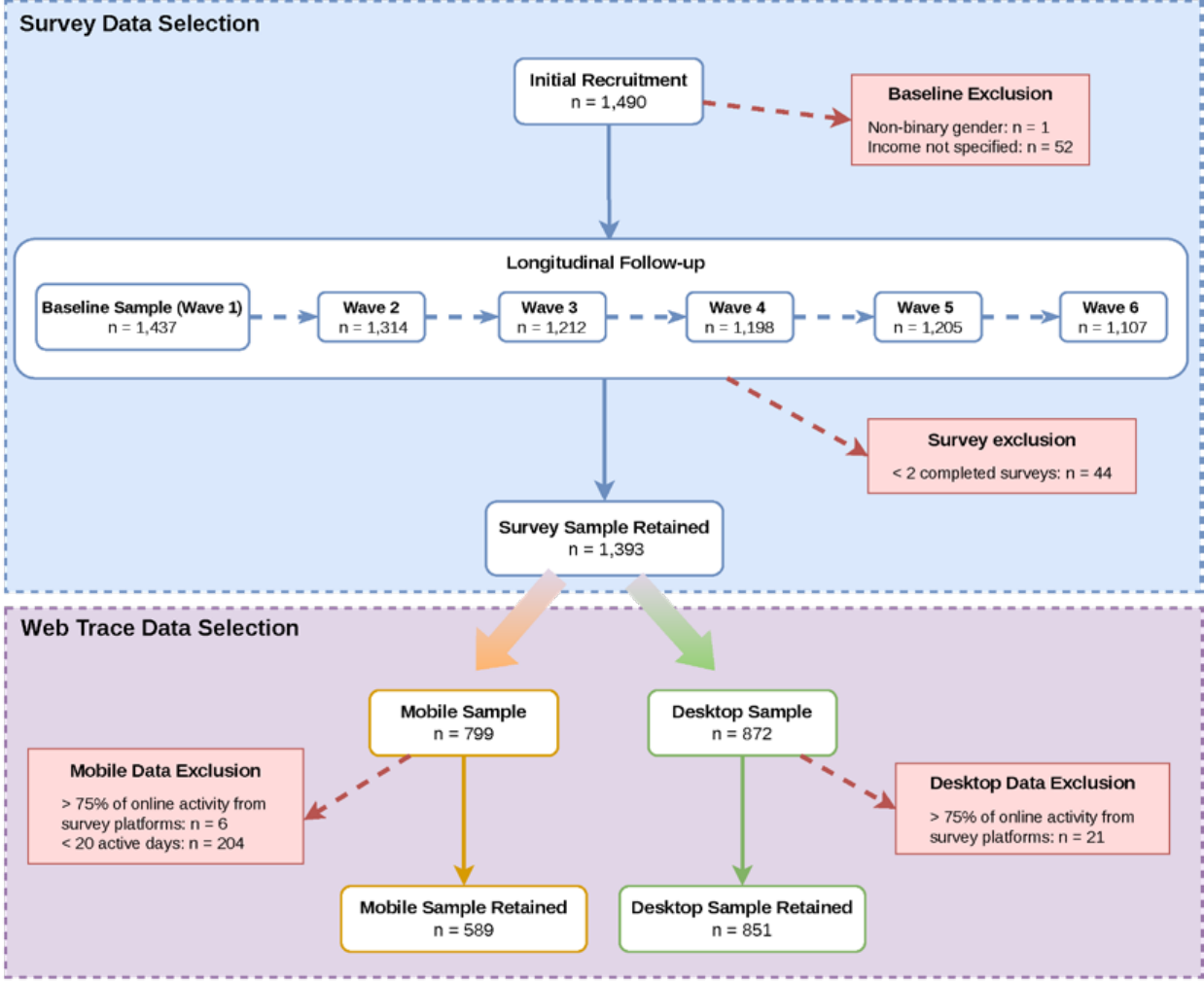}
    \caption{Participant recruitment and exclusion process. Participants could contribute data to both analyses if they met the quality criteria for both devices.}
    \label{fig:flow}
\end{figure}

To ensure data quality, we applied systematic filtering criteria to the web trace data. 
We excluded panelists who appeared to be professional survey-takers, defined as those spending $\geq{75}\%$ of their logged time on survey websites. 
For mobile users, we required a minimum of 20 active days within each four-week measurement window corresponding to survey waves. 
These criteria strike a balance between excluding participants with insufficient web trace data and preserving diversity in digital usage patterns. 
After applying these filters, our final analytic sample comprised 589 mobile users and 851 desktop users across all study waves (Figure~\ref{fig:flow}).

All participants provided informed consent for both the questionnaire and the web trace data collection. The study received approval from the University's Research Ethics Committee.

\subsection{Quantifying social media use}\label{sec:quantSM}
We quantified social media use by transforming raw web traces into interpretable behavior features. We defined social media platforms as online services designed for communication, social interaction, and content sharing, encompassing messaging services, dating applications, video-sharing platforms, and online forums. 
This definition covered 107 domains in the desktop dataset and 73 applications and domains in the mobile dataset. 
For each participant and survey wave, we calculated the average daily time spent on social media during the four weeks preceding survey completion, expressed in hours per day. 
This approach provided a stable measure of social media exposure that aligned with our monthly survey schedule.

\textbf{Temporal engagement patterns.} We constructed temporal features to capture participants' engagement patterns and behavioral rhythms. 
Individual usage records were aggregated into sessions, with consecutive activities separated by less than 30 minutes merged to account for brief interruptions. 
From these sessions, we generated four temporal metrics: \textbf{Daily Sessions} measured the average number of distinct usage sessions per day; \textbf{Session Duration} captured the average length of each session in hours; \textbf{Inter-session Interval} calculated the average time gap between consecutive sessions in hours; and \textbf{Work-Free Difference} quantified the total difference in social media time between free hours and work hours (defined as 9 AM to 5 PM on weekdays), summed across the four-week measurement period and expressed in total hours.

\textbf{Platform categorization.} To examine whether different platform types were differentially associated with loneliness, we classified social media applications and domains used by at least 100 participants ($n = 29$ platforms, see Appendix~\ref{secA5} for all platform categories). 
Classification was based on official app store descriptions and platform functionality. 
The first author conducted the initial categorization, which was then reviewed and validated by the other two co-authors. Any disagreements were resolved through discussion to reach a consensus. 
We identified five categories: (a) \textbf{messaging services} (WhatsApp, Telegram); (b) \textbf{visual-sharing platforms }(Instagram, Pinterest, TikTok, YouTube); (c) \textbf{networking-oriented platforms }(Facebook, LinkedIn, Twitter/X); (d) \textbf{relationship-oriented platforms }(dating and friendship applications); and (e) \textbf{discussion forums} (community-based platforms like Reddit).

This feature construction allowed us to test our three hypotheses examining the association between loneliness and (H1) social media time, (H2) temporal engagement patterns, and (H3) platform type usage behaviors.

\subsection{Measuring loneliness}
Loneliness was assessed monthly using the three-item UCLA Loneliness Scale (UCLA-3)~\cite{Hughes2004ShortUCLA}. The scale includes items such as ``How often do you feel isolated from others?'', with response options ranging from 1 (hardly ever) to 3 (often). 
Scores were summed to create a total score ranging from 3 to 9, with higher values indicating greater loneliness. 
The scale demonstrated high internal consistency in our sample (Cronbach’s $\alpha = 0.82$, 95\% CI [0.82, 0.83]), consistent with established psychometric evidence for the UCLA-3 in large-scale studies~\cite{Hughes2004ShortUCLA}. 

\textbf{Analytic approach.} We used UCLA-3 scores as continuous variables in our primary analyses (H1). For hypotheses examining group differences (H2, H3), we created comparison groups based on participants' average loneliness scores across all available waves. 
Different methods have been used in prior research to determine loneliness status, including cut-off scores based on distributional percentiles or absolute thresholds such as scores exceeding 5, 6, or 7~\cite{steptoe2013social,das2021systematic,lasgaard2016where,matthews2022using}. Following established evidence that a cut-off of 5 effectively identifies individuals at elevated risk for mental health problems~\cite{matthews2022using}, participants with average scores $>5$ were classified as ``lonely'' ($n = 160$ mobile, $n = 225$ desktop) and those with average scores of 3 were classified as ``non-lonely'' ($n = 131$ mobile, $n = 213$ desktop). Participants with intermediate scores were excluded from group-based analyses to create distinct comparison groups that maximized between-group differences.

\subsection{Statistical analysis}\label{sec:statanalysis}


We analyzed the data using linear mixed-effects models (LMMs) implemented in R (lme4 package)~\cite{lme4}, with random intercepts for participants to account for the repeated measures. In the model specification, $i$ denotes participants and $t$ denotes waves. The general model specification was:
\begin{equation}
    \mathrm{UCLA} 3_{i t}=\beta_{0}+\mathbf{x}_{i t}^{\prime} \boldsymbol{\beta}+u_{i}+\varepsilon_{i t}, \quad u_{i} \sim \mathcal{N}\left(0, \tau^{2}\right), \quad \varepsilon_{i t} \sim \mathcal{N}\left(0, \sigma^{2}\right),
\end{equation}

where $x_{it}$ represents the fixed effects of interest. Models were estimated separately for mobile and desktop datasets. Fixed effects were tested using Satterthwaite’s approximation for degrees of freedom (R package lmerTest~\cite{Kuznetsova2017lmerTest}). We assessed multicollinearity using dimension-adjusted generalized variance inflation factors (GVIF), an extension of variance inflation factors for categorical predictors~\cite{Fox1992GVIF}. All adjusted GVIF values were below 2, and model fit was evaluated using marginal and conditional $R^2$. The intraclass correlation coefficient (ICC) quantified the proportion of variance attributable to between-person differences. All categorical predictors were dummy-coded with the following reference categories: male (gender), 18–30 (age group), $\leq3000$\texteuro (income group), single (relationship status), and intermediate (chronotype).

\textbf{H1: Social media and loneliness association.} 
We estimated LMMs that included sociodemographic covariates (wave, gender, age group, income group, relationship status, and chronotype), total time spent online, and social media use (Model M1). To validate robustness, we extended this model by adding all online activity categories (entertainment, games, news, search, shopping, social media, productivity; Model M2). 
We conducted subgroup analyses stratified by demographic characteristics to examine consistency across population groups.

\textbf{H2 and H3: Group comparisons. }
For temporal patterns (H2) and platform categories (H3), we compared participants classified as lonely ($score > 5$) with those classified as non-lonely ($score = 3$). Because intra-individual variation was minimal across waves, we aggregated data at the participant level using mean values to ensure independent observations. 
We applied z-score-based outlier detection and replacement within each group separately. Values exceeding three standard deviations from the group mean were replaced with the group median to reduce the influence of extreme outliers while preserving sample size. 
Group differences were evaluated using Welch's t-tests (robust to unequal variances). Effect sizes were calculated as Hedges' g, an unbiased estimator that applies a correction factor to Cohen's d to account for small sample bias. We report Hedges' g with 90\% confidence intervals estimated via bias-corrected bootstrap resampling (5,000 iterations).
We conducted two one-sided tests (TOST) for equivalence testing against equivalence bounds defined by a smallest effect size of interest (SESOI) of 0.4 (Hedges’ g). 
This threshold approximates a medium effect size~\cite{cohen1988power} and has been applied in recent research on social media use and mental health outcomes~\cite{Fassi2025MHCYP}. 
Following recommendations for theoretically and practically grounded SESOI selection~\cite{lakens2018equivalence}, we chose this value to balance sensitivity for detecting meaningful differences while avoiding overinterpretation of trivial effects. 
For each comparison, the SESOI was converted into raw units using the pooled standard deviation. 
This approach distinguished between practically meaningful differences, inconclusive results, and negligible effects. P-values were adjusted using the Benjamini–Hochberg procedure for false discovery rate control (FDR-BH), with significance set at $\alpha=0.05$.

\subsection{Sensitivity Analyses}
To assess the robustness of our findings to methodological choices, we conducted extensive sensitivity analyses examining key analytical decisions and data characteristics.

\textbf{Data quality filtering thresholds.} We assessed whether our participant selection criteria influenced main findings by testing alternative thresholds for professional survey-taker exclusion (70\% and 80\% vs. our primary 75\% threshold) and mobile activity requirements (15 and 25 days vs. our primary 20-day threshold). For each alternative threshold, we re-ran the primary social media-loneliness association model (Model M1) to examine consistency of regression coefficients and statistical significance. 

\textbf{Preprocessing method comparison.} We compared seven different preprocessing approaches: (1) raw data with no preprocessing, (2) IQR-based winsorization with varying multipliers (k = 1.0, 1.5, 2.0), (3) log1p transformation, (4) square root transformation, and (5) z-score-based outlier removal (threshold = 3 SD with median replacement). Each preprocessing method was applied within groups separately, and identical statistical testing procedures (Welch's t-tests, Mann-Whitney U tests, TOST equivalence tests, and FDR-BH correction) were used to enable direct comparison of results. 

\textbf{Zero-inflated feature identification and analysis.} We identified features with high proportions of zero values (threshold $> 50\%$ in either group) as zero-inflated and conducted supplementary analyses specifically designed for such data. 
For zero-inflated features, we performed: (1) usage rate analysis comparing the proportion of users with non-zero values between groups using proportion z-tests and Cohen's h effect sizes, and (2) conditional analysis of non-zero values using Mann-Whitney U tests among users who actually engaged with the platform type.

\textbf{Statistical method robustness.} We compared parametric (Welch's t-test) and non-parametric (Mann-Whitney U test) approaches across all preprocessing methods to assess sensitivity to distributional assumptions. 
Agreement between methods was quantified, and we examined cases where conclusions differed between parametric and non-parametric tests.

\textbf{Method stability assessment.} We evaluated the consistency of effect sizes across preprocessing methods by calculating the average deviation of each method's effect sizes from the overall mean effect size per feature. This provided a ranking of preprocessing methods based on their stability and robustness to analytical choices.
For each sensitivity analysis, we examined changes in: (1) the number and identity of statistically significant effects after FDR correction, (2) effect size magnitudes (Hedges' g and Cohen's h), (3) the practical interpretation of findings based on equivalence testing results, and (4) agreement between different statistical approaches. 

\section{Declaration of generative AI and AI-assisted technologies in the writing process}

During the preparation of this work the authors used ChatGPT for text proof reading (including spelling and grammar checks). After using this tool, the authors carefully reviewed and edited the content as necessary and take full responsibility for the final content of the published article.

\clearpage
\bibliographystyle{plain}
\bibliography{ref}

\appendix
\clearpage
\begin{appendices}

\section{Sample Characteristics}\label{secA4}
\begin{table}[htbp]
\centering
\caption{Demographic characteristics of the study sample across mobile and desktop users. Values are reported as the number of participants (percentage).}
\label{tab:demographics}
\begin{tabular}{lll}
\hline
\textbf{Demographic Category} & \textbf{Mobile n (\%)} & \textbf{Desktop n (\%)} \\
\hline
\textbf{Income} & & \\
\quad $\leq$ 3000        & 326 (55.3\%) & 481 (56.5\%) \\
\quad $\geq$ 3001        & 263 (44.7\%) & 370 (43.5\%) \\
\textbf{Gender} & & \\
\quad Male               & 318 (53.9\%) & 461 (54.2\%) \\
\quad Female             & 271 (46.1\%) & 390 (45.8\%) \\
\textbf{Age group} & & \\
\quad 18--30             & 37 (6.3\%)   & 67 (7.9\%)   \\
\quad 31--60             & 443 (75.2\%) & 605 (71.1\%) \\
\quad 60+                & 109 (18.5\%) & 179 (21.0\%) \\
\textbf{Relationship status} & & \\
\quad Single             & 175 (29.7\%) & 268 (31.5\%) \\
\quad With partner       & 283 (48.0\%) & 409 (48.1\%) \\
\textbf{Chronotype} & & \\
\quad Intermediate        & 305 (51.8\%) & 412 (48.4\%) \\
\quad Morning type         & 222 (37.7\%) & 347 (40.8\%) \\
\quad Evening type         & 62 (10.5\%)  & 92 (10.8\%)  \\
\hline
\end{tabular}
\end{table}

\clearpage
\section{LLMs Results(H1)}\label{secA1}

\begin{table}[htbp]
\centering
\caption{Linear mixed-effects model predicting UCLA-3 loneliness scores from \textbf{mobile usage data}.
The model includes fixed effects for demographic and behavioral variables, and a random intercept for the participant.
Estimates are unstandardized coefficients.}
\label{tab:lmm-mobile}
\begin{tabular}{lrrrrr}
\toprule
Predictor & Estimate & Std.\ Error & df & $t$ & $p$ \\
\midrule
Intercept                  & 5.258 & 0.273 &  650 & 19.265 & $<0.001$ \\
Wave                       & -0.034 & 0.010 & 2417 & -3.578 & $<0.001$ \\
Female (vs.\ Male)         & 0.159 & 0.117 &  580 &  1.364 & 0.173 \\
Age 31--60 (vs.\ 18--30)   & -0.327 & 0.244 &  598 & -1.339 & 0.181 \\
Age 60+ (vs.\ 18--30)      & -0.292 & 0.274 &  598 & -1.066 & 0.287 \\
Income $>3000$ (vs.\ $\leq3000$) & -0.244 & 0.127 &  581 & -1.920 & 0.055 \\
With partner (vs.\ single) & -0.595 & 0.146 &  563 & -4.080 & $<0.001$ \\
Relationship unknown (vs.\ single) & 0.015 & 0.169 & 612 & 0.091 & 0.928 \\
Morning type (vs.\ Intermediate) & -0.458 & 0.124 & 576 & -3.699 & $<0.001$ \\
Evening type (vs.\ Intermediate) & -0.136 & 0.195 & 576 & -0.695 & 0.487 \\
Social media time          & 0.085 & 0.040 & 2496 & 2.121 & 0.034 \\
Total screen time          & -0.017 & 0.020 & 2390 & -0.844 & 0.399 \\
\bottomrule
\end{tabular}
\vspace{2pt}
\begin{minipage}{0.96\linewidth}
\footnotetext{Random effects: variance of random intercept (participant) $= 1.750$, $SD = 1.323$; residual variance $= 0.763$, $SD = 0.873$. 
Model fit: $AIC = 9155.8$; $BIC = 9239.8$; $\mathrm{logLik} = -4563.92$. 
ICC: $0.696$. 
Explained variance: marginal $R^2 = 0.086$; conditional $R^2 = 0.723$.}
\end{minipage}
\end{table}

\begin{table}[htbp]
\centering
\caption{Linear mixed-effects model predicting UCLA-3 loneliness scores from \textbf{desktop usage data}.
The model includes fixed effects for demographic and behavioral variables, and a random intercept for the participant. 
Estimates are unstandardized coefficients.} 
\label{tab:lmm-desktop}
\begin{tabular}{lrrrrr}
\toprule
Predictor & Estimate & Std.\ Error & df & $t$ & $p$ \\
\midrule
Intercept                  & 5.285 & 0.206 & 915  & 25.692 & $<0.001$ \\
Wave                       & -0.011 & 0.009 & 2721 & -1.130 & 0.259 \\
Female (vs.\ Male)         & 0.083 & 0.097 & 837  &  0.853 & 0.394 \\
Age 31--60 (vs.\ 18--30)   & -0.404 & 0.183 & 880  & -2.202 & 0.028 \\
Age 60+ (vs.\ 18--30)      & -0.629 & 0.205 & 874  & -3.073 & 0.002 \\
Income $>3000$ (vs.\ $\leq3000$) & -0.383 & 0.106 & 837  & -3.614 & $<0.001$ \\
With partner (vs.\ single) & -0.464 & 0.120 & 821  & -3.874 & $<0.001$ \\
Relationship unknown (vs.\ single) & 0.074 & 0.142 & 887 & 0.517 & 0.606 \\
Morning type (vs.\ Intermediate) & -0.410 & 0.102 & 843 & -4.019 & $<0.001$ \\
Evening type (vs.\ Intermediate) & -0.069 & 0.161 & 853 & -0.431 & 0.667 \\
Social media time          & 0.102 & 0.043 & 2973 & 2.403 & 0.016 \\
Total screen time          & -0.029 & 0.018 & 3155 & -1.594 & 0.111 \\
\bottomrule
\end{tabular}
\vspace{2pt}
\begin{minipage}{0.96\linewidth}
\footnotetext{Random effects: variance of random intercept (participant) $= 1.643$, $SD = 1.282$; residual variance $= 0.785$, $SD = 0.886$. 
Model fit: $AIC = 10{,}744.7$; $BIC = 10{,}830.6$; $\mathrm{logLik} = -5358.36$. 
ICC: $0.677$. 
Explained variance: marginal $R^2 = 0.087$; conditional $R^2 = 0.705$.}
\end{minipage}
\end{table}

\begin{table}[htbp]
\centering
\caption{Linear mixed-effects model predicting UCLA-3 loneliness scores from \textbf{mobile usage data with activity categories}.
The model includes fixed effects for demographic and behavioral variables, as well as a random intercept for each participant.
Estimates are unstandardized coefficients.}
\label{tab:lmm-mobile-activities}
\begin{tabular}{lrrrrr}
\toprule
Predictor & Estimate & Std.\ Error & df & $t$ & $p$ \\
\midrule
Intercept                  & 5.279 & 0.275 &  657 & 19.184 & $<0.001$ \\
Wave                       & -0.033 & 0.010 & 2439 & -3.398 & $<0.001$ \\
Female (vs.\ Male)         & 0.161 & 0.118 &  595 &  1.362 & 0.174 \\
Age 31--60 (vs.\ 18--30)   & -0.324 & 0.246 &  601 & -1.319 & 0.188 \\
Age 60+ (vs.\ 18--30)      & -0.279 & 0.275 &  602 & -1.013 & 0.312 \\
Income $>3000$ (vs.\ $\leq3000$) & -0.239 & 0.128 &  585 & -1.868 & 0.062 \\
With partner (vs.\ single) & -0.591 & 0.147 &  565 & -4.036 & $<0.001$ \\
Relationship unknown (vs.\ single) & 0.033 & 0.170 &  612 &  0.192 & 0.848 \\
Morning type (vs.\ Intermediate) & -0.451 & 0.124 &  577 & -3.621 & $<0.001$ \\
Evening type (vs.\ Intermediate) & -0.116 & 0.196 &  577 & -0.592 & 0.554 \\
Entertainment              & 0.004 & 0.003 & 2919 &  1.213 & 0.225 \\
Games                      & 0.001 & 0.002 & 2902 &  0.626 & 0.532 \\
Health-related                     & 0.002 & 0.013 & 2931 &  0.140 & 0.889 \\
News                       & -0.001 & 0.007 & 2657 & -0.097 & 0.923 \\
Search                     & -0.005 & 0.010 & 2873 & -0.533 & 0.594 \\
Shopping                   & -0.005 & 0.005 & 2712 & -0.998 & 0.318 \\
Social media total time              & 0.003 & 0.002 & 2889 &  1.739 & 0.082 \\
Productivity               & 0.001 & 0.004 & 2958 &  0.328 & 0.743 \\
Total screen time          & -0.032 & 0.042 & 2911 & -0.780 & 0.436 \\
\bottomrule
\end{tabular}
\vspace{2pt}
\begin{minipage}{0.96\linewidth}
\footnotetext{Random effects: variance of random intercept (participant) $= 1.760$, $SD = 1.327$; residual variance $= 0.763$, $SD = 0.874$. 
Model fit: $AIC = 9234.4$; $BIC = 9360.4$; $\mathrm{logLik} = -4596.22$. 
ICC: $0.698$. 
Explained variance: marginal $R^2 = 0.087$; conditional $R^2 = 0.724$. 
All activity predictors (entertainment, games, health, news, search, shopping, social media, and productivity) are measured in terms of average daily hours.}
\end{minipage}
\end{table}

\begin{table}[htbp]
\centering
\caption{Linear mixed-effects model predicting UCLA-3 loneliness scores from \textbf{desktop usage data with activity categories}.
The model includes fixed effects for demographic and behavioral variables, and a random intercept for the participant.
Estimates are unstandardized coefficients.}
\label{tab:lmm-desktop-activities}
\begin{tabular}{lrrrrr}
\toprule
Predictor & Estimate & Std.\ Error & df & $t$ & $p$ \\
\midrule
Intercept                  & 5.306 & 0.206 &  922 & 25.748 & $<0.001$ \\
Wave                       & -0.012 & 0.010 & 2721 & -1.241 & 0.215 \\
Female (vs.\ Male)         & 0.074 & 0.097 &  844 &  0.763 & 0.446 \\
Age 31--60 (vs.\ 18--30)   & -0.403 & 0.184 &  881 & -2.197 & 0.028 \\
Age 60+ (vs.\ 18--30)      & -0.616 & 0.205 &  879 & -3.003 & 0.003 \\
Income $>3000$ (vs.\ $\leq3000$) & -0.384 & 0.106 &  838 & -3.612 & $<0.001$ \\
With partner (vs.\ single) & -0.467 & 0.120 &  827 & -3.885 & $<0.001$ \\
Relationship unknown (vs.\ single) & 0.072 & 0.143 &  889 & 0.502 & 0.616 \\
Morning type (vs.\ Intermediate) & -0.414 & 0.102 &  845 & -4.044 & $<0.001$ \\
Evening type (vs.\ Intermediate) & -0.073 & 0.161 &  854 & -0.454 & 0.650 \\
Entertainment              & -0.002 & 0.002 & 3296 & -1.193 & 0.233 \\
Games                      & -0.004 & 0.004 & 3052 & -0.995 & 0.320 \\
Health-related                     &  0.047 & 0.088 & 2909 &  0.537 & 0.592 \\
News                       & -0.008 & 0.005 & 3388 & -1.622 & 0.105 \\
Search                     &  0.000 & 0.003 & 2267 &  0.049 & 0.961 \\
Shopping                   &  0.004 & 0.003 & 3298 &  1.338 & 0.181 \\
Social media total time               &  0.004 & 0.002 & 3187 &  2.472 & 0.013 \\
Work-related               & -0.004 & 0.004 & 3382 & -1.004 & 0.316 \\
Total screen time          & -0.018 & 0.024 & 3396 & -0.744 & 0.457 \\
\bottomrule
\end{tabular}
\vspace{2pt}
\begin{minipage}{0.96\linewidth}
\footnotetext{Random effects: variance of random intercept (participant) $= 1.645$, $SD = 1.283$; residual variance $= 0.785$, $SD = 0.886$. 
Model fit: $AIC = 10{,}817.7$; $BIC = 10{,}946.6$; $\mathrm{logLik} = -5387.84$. 
ICC: $0.677$. 
Explained variance: marginal $R^2 = 0.090$; conditional $R^2 = 0.706$. 
All activity predictors (entertainment, games, health, news, search, shopping, social media, and productivity) are measured in terms of average daily hours.}
\end{minipage}
\end{table}

\clearpage

\subsection{Demographic features subgroup LLM Results}\label{secA2}

\begin{table}[htbp]
\centering
\caption{Summary of \textbf{social media effects} on UCLA-3 loneliness scores across demographic and chronotype subgroups, 
estimated separately for mobile and desktop data. 
Estimates are unstandardized coefficients ($b$) with standard errors (SE). 
Significance is based on Satterthwaite $t$-tests.}
\label{tab:subgroup-summary-socialmedia}
\begin{tabular}{lrrrr}
\toprule
 & \multicolumn{2}{c}{Mobile} & \multicolumn{2}{c}{Desktop} \\
\cmidrule(lr){2-3} \cmidrule(lr){4-5}
Subgroup & $b$ (SE) & $p$ & $b$ (SE) & $p$ \\
\midrule
Income $\leq 3000$   & 0.076 (0.050) & 0.134 & 0.081 (0.055) & 0.144 \\
Income $> 3000$      & 0.101 (0.066) & 0.127 & 0.141 (0.069) & \textbf{0.039*} \\
Male                 & 0.038 (0.056) & 0.495 & 0.083 (0.066) & 0.211 \\
Female               & 0.133 (0.058) & \textbf{0.021*} & 0.119 (0.057) & \textbf{0.038*} \\
Single               & 0.088 (0.070) & 0.211 & 0.110 (0.070) & 0.116 \\
With partner         & 0.099 (0.055) & 0.075 & 0.111 (0.065) & 0.087 \\
Relationship unknown & 0.051 (0.105) & 0.626 & 0.060 (0.115) & 0.599 \\
Intermediate chronotype   & 0.094 (0.057) & 0.100 & 0.190 (0.058) & \textbf{0.001**} \\
Morning type   & 0.146 (0.063) & \textbf{0.021*} & -0.021 (0.077) & 0.783 \\
Evening type  & 0.032 (0.125) & 0.800 & -0.119 (0.116) & 0.307 \\
Age 18--30           & -0.128 (0.199) & 0.520 & -0.285 (0.221) & 0.197 \\
Age 31--60           &  0.123 (0.045) & \textbf{0.006**} & 0.083 (0.052) & 0.110 \\
Age 60+              & -0.112 (0.109) & 0.306 & 0.207 (0.082) & \textbf{0.011*} \\
\bottomrule
\end{tabular}
\end{table}

\begin{table}[htbp]
\centering
\caption{Subgroup analysis: \textbf{Income $\leq 3000$ (Mobile)}. 
Linear mixed-effects model predicting UCLA-3 loneliness scores among participants with low income.
Estimates are unstandardized coefficients.}
\label{tab:subgroup-income-low-mobile}
\begin{tabular}{lrrrrr}
\toprule
Predictor & Estimate & Std.\ Error & df & $t$ & $p$ \\
\midrule
Intercept                     & 5.107 & 0.376 &   354 & 13.573 & $<0.001$ \\
Wave                          & -0.039 & 0.013 &  1350 & -3.027 & 0.003 \\
Female (vs.\ Male)            &  0.214 & 0.167 &   322 &  1.281 & 0.201 \\
Age 31--60 (vs.\ 18--30)      & -0.269 & 0.353 &   325 & -0.763 & 0.446 \\
Age 60+ (vs.\ 18--30)         &  0.018 & 0.384 &   326 &  0.048 & 0.962 \\
With partner (vs.\ Single)    & -0.502 & 0.189 &   311 & -2.665 & 0.008 \\
Relationship unknown (vs.\ Single) & 0.245 & 0.221 &   340 &  1.105 & 0.270 \\
Morning type (vs.\ Intermediate)   & -0.560 & 0.182 &   318 & -3.083 & 0.002 \\
Evening type (vs.\ Intermediate)   & -0.530 & 0.264 &   315 & -2.010 & 0.045 \\
Social media time             &  0.076 & 0.050 &  1491 &  1.501 & 0.134 \\
Total screen time             & -0.004 & 0.026 &  1457 & -0.165 & 0.869 \\
\bottomrule
\end{tabular}
\vspace{2pt}
\begin{minipage}{0.96\linewidth}
\footnotetext{Sample size: 1661 observations from 326 participants. \\
Random effects: variance of random intercept (participant) $= 2.028$, $SD = 1.424$; residual variance $= 0.752$, $SD = 0.867$. \\
Model fit: $AIC = 5141.9$; $BIC = 5212.3$; $\mathrm{logLik} = -2558.0$. \\
Intraclass correlation: $ICC = 0.730$. \\
Explained variance: marginal $R^2 = 0.072$; conditional $R^2 = 0.749$.}
\end{minipage}
\end{table}

\begin{table}[htbp]
\centering
\caption{Subgroup analysis: \textbf{Income $\leq 3000$ (Desktop)}. 
Linear mixed-effects model predicting UCLA-3 loneliness scores among low-income participants using desktop data.
Estimates are unstandardized coefficients.}
\label{tab:subgroup-income0-desktop}
\begin{tabular}{lrrrrr}
\toprule
Predictor & Estimate & Std.\ Error & df & $t$ & $p$ \\
\midrule
Intercept                  & 5.267 & 0.301 &   511 & 17.477 & $<0.001$ \\
Wave                       &  0.008 & 0.013 &  1542 &  0.591 & 0.554 \\
Female (vs.\ Male)         &  0.131 & 0.141 &   468 &  0.930 & 0.353 \\
Age 31--60 (vs.\ 18--30)   & -0.492 & 0.275 &   492 & -1.789 & 0.074 \\
Age 60+ (vs.\ 18--30)      & -0.519 & 0.298 &   490 & -1.741 & 0.082 \\
With partner (vs.\ Single) & -0.536 & 0.158 &   458 & -3.400 & $<0.001$ \\
Relationship unknown (vs.\ Single) & 0.154 & 0.194 &   507 &  0.792 & 0.429 \\
Morning type (vs.\ Intermediate) & -0.452 & 0.152 &   476 & -2.970 & 0.003 \\
Evening type (vs.\ Intermediate) & -0.228 & 0.222 &   482 & -1.025 & 0.306 \\
Social media time          &  0.081 & 0.055 &  1660 &  1.461 & 0.144 \\
Total screen time          & -0.018 & 0.024 &  1786 & -0.741 & 0.459 \\
\bottomrule
\end{tabular}
\vspace{2pt}
\begin{minipage}{0.96\linewidth}
\footnotetext{Sample size: 1948 observations from 481 participants. \\
Random effects: variance of random intercept (participant) $= 2.016$, $SD = 1.420$; residual variance $= 0.890$, $SD = 0.943$. \\
Model fit: $AIC = 6410.6$; $BIC = 6483.0$; $\mathrm{logLik} = -3192.3$. \\
Intraclass correlation: $ICC = 0.694$. \\
Explained variance: marginal $R^2 = 0.054$; conditional $R^2 = 0.711$.}
\end{minipage}
\end{table}

\begin{table}[htbp]
\centering
\caption{Subgroup analysis: \textbf{Income $>$ 3000 (Mobile)}. 
Linear mixed-effects model predicting UCLA-3 loneliness scores among participants with high income.
Estimates are unstandardized coefficients.}
\label{tab:subgroup-income-high-mobile}
\begin{tabular}{lrrrrr}
\toprule
Predictor & Estimate & Std.\ Error & df & $t$ & $p$ \\
\midrule
Intercept                     & 5.553 & 0.405 &   275 & 13.718 & $<0.001$ \\
Wave                          & -0.029 & 0.015 &  1062 & -2.020 & 0.044 \\
Female (vs.\ Male)            & -0.005 & 0.156 &   249 & -0.033 & 0.974 \\
Age 31--60 (vs.\ 18--30)      & -0.421 & 0.324 &   264 & -1.299 & 0.195 \\
Age 60+ (vs.\ 18--30)         & -0.907 & 0.379 &   263 & -2.391 & 0.018 \\
With partner (vs.\ Single)    & -1.048 & 0.251 &   236 & -4.174 & $<0.001$ \\
Relationship unknown (vs.\ Single) & -0.653 & 0.286 &   254 & -2.285 & 0.023 \\
Morning type (vs.\ Intermediate)   & -0.239 & 0.162 &   247 & -1.470 & 0.143 \\
Evening type (vs.\ Intermediate)   &  0.593 & 0.284 &   255 &  2.089 & 0.038 \\
Social media time             &  0.101 & 0.066 &   919 &  1.526 & 0.127 \\
Total screen time             & -0.031 & 0.033 &   837 & -0.945 & 0.345 \\
\bottomrule
\end{tabular}
\vspace{2pt}
\begin{minipage}{0.96\linewidth}
\footnotetext{Sample size: 1317 observations from 263 participants. \\
Random effects: variance of random intercept (participant) $= 1.311$, $SD = 1.145$; residual variance $= 0.777$, $SD = 0.882$. \\
Model fit: $AIC = 4023.3$; $BIC = 4090.6$; $\mathrm{logLik} = -1998.6$. \\
Intraclass correlation: $ICC = 0.628$. \\
Explained variance: marginal $R^2 = 0.114$; conditional $R^2 = 0.670$.}
\end{minipage}
\end{table}

\begin{table}[htbp]
\centering
\caption{Subgroup analysis: \textbf{Income $>$ 3000 (Desktop)}. 
Linear mixed-effects model predicting UCLA-3 loneliness scores among higher-income participants using desktop data.
Estimates are unstandardized coefficients.}
\label{tab:subgroup-income1-desktop}
\begin{tabular}{lrrrrr}
\toprule
Predictor & Estimate & Std.\ Error & df & $t$ & $p$ \\
\midrule
Intercept                  & 4.916 & 0.286 &   382 & 17.185 & $<0.001$ \\
Wave                       & -0.034 & 0.013 &  1181 & -2.569 & 0.010 \\
Female (vs.\ Male)         & -0.019 & 0.126 &   359 & -0.150 & 0.881 \\
Age 31--60 (vs.\ 18--30)   & -0.318 & 0.231 &   377 & -1.376 & 0.170 \\
Age 60+ (vs.\ 18--30)      & -0.878 & 0.269 &   374 & -3.269 & 0.001 \\
With partner (vs.\ Single) & -0.373 & 0.208 &   348 & -1.796 & 0.073 \\
Relationship unknown (vs.\ Single) & 0.020 & 0.234 &   361 &  0.085 & 0.932 \\
Morning type (vs.\ Intermediate) & -0.344 & 0.129 &   360 & -2.666 & 0.008 \\
Evening type (vs.\ Intermediate) & 0.278 & 0.229 &   363 &  1.215 & 0.225 \\
Social media time          &  0.141 & 0.069 &  1335 &  2.063 & 0.039 \\
Total screen time          & -0.053 & 0.029 &  1382 & -1.829 & 0.067 \\
\bottomrule
\end{tabular}
\vspace{2pt}
\begin{minipage}{0.96\linewidth}
\footnotetext{Sample size: 1470 observations from 370 participants. \\
Random effects: variance of random intercept (participant) $= 1.151$, $SD = 1.073$; residual variance $= 0.644$, $SD = 0.802$. \\
Model fit: $AIC = 4304.4$; $BIC = 4373.2$; $\mathrm{logLik} = -2139.2$. \\
Intraclass correlation: $ICC = 0.641$. \\
Explained variance: marginal $R^2 = 0.092$; conditional $R^2 = 0.674$.}
\end{minipage}
\end{table}

\begin{table}[htbp]
\centering
\caption{Subgroup analysis: \textbf{Male participants (Mobile)}. 
Linear mixed-effects model predicting UCLA-3 loneliness scores among male participants.
Estimates are unstandardized coefficients.}
\label{tab:subgroup-gender-male-mobile}
\begin{tabular}{lrrrrr}
\toprule
Predictor & Estimate & Std.\ Error & df & $t$ & $p$ \\
\midrule
Intercept                     & 5.372 & 0.407 &   340 & 13.197 & $<0.001$ \\
Wave                          & -0.028 & 0.012 &  1318 & -2.246 & 0.025 \\
Income $>$3000 (vs.\ $\leq$3000) & -0.126 & 0.169 &   306 & -0.743 & 0.458 \\
Age 31--60 (vs.\ 18--30)      & -0.279 & 0.392 &   320 & -0.712 & 0.477 \\
Age 60+ (vs.\ 18--30)         & -0.258 & 0.431 &   318 & -0.597 & 0.551 \\
With partner (vs.\ Single)    & -0.738 & 0.198 &   300 & -3.718 & $<0.001$ \\
Relationship unknown (vs.\ Single) & -0.134 & 0.241 &   327 & -0.556 & 0.578 \\
Morning type (vs.\ Intermediate)   & -0.588 & 0.167 &   305 & -3.524 & $<0.001$ \\
Evening type (vs.\ Intermediate)   & -0.397 & 0.285 &   304 & -1.393 & 0.165 \\
Social media time             &  0.038 & 0.056 &  1406 &  0.682 & 0.495 \\
Total screen time             & -0.022 & 0.028 &  1439 & -0.777 & 0.437 \\
\bottomrule
\end{tabular}
\vspace{2pt}
\begin{minipage}{0.96\linewidth}
\footnotetext{Sample size: 1618 observations from 318 participants. \\
Random effects: variance of random intercept (participant) $= 1.733$, $SD = 1.316$; residual variance $= 0.690$, $SD = 0.831$. \\
Model fit: $AIC = 4848.5$; $BIC = 4918.5$; $\mathrm{logLik} = -2411.2$. \\
Intraclass correlation: $ICC = 0.715$. \\
Explained variance: marginal $R^2 = 0.091$; conditional $R^2 = 0.741$.}
\end{minipage}
\end{table}

\begin{table}[htbp]
\centering
\caption{Subgroup analysis: \textbf{Male participants (Desktop)}. 
Linear mixed-effects model predicting UCLA-3 loneliness scores among male participants using desktop data.
Estimates are unstandardized coefficients.}
\label{tab:subgroup-gender0-desktop}
\begin{tabular}{lrrrrr}
\toprule
Predictor & Estimate & Std.\ Error & df & $t$ & $p$ \\
\midrule
Intercept                  & 5.363 & 0.313 &   497 & 17.146 & $<0.001$ \\
Wave                       & -0.035 & 0.012 &  1529 & -2.795 & 0.005 \\
Income $>$3000 (vs.\ $\leq$3000) & -0.246 & 0.139 &   456 & -1.773 & 0.077 \\
Age 31--60 (vs.\ 18--30)   & -0.307 & 0.302 &   471 & -1.017 & 0.310 \\
Age 60+ (vs.\ 18--30)      & -0.556 & 0.331 &   468 & -1.681 & 0.093 \\
With partner (vs.\ Single) & -0.644 & 0.159 &   450 & -4.051 & $<0.001$ \\
Relationship unknown (vs.\ Single) & -0.155 & 0.200 &   485 & -0.773 & 0.440 \\
Morning type (vs.\ Intermediate) & -0.427 & 0.136 &   463 & -3.149 & 0.002 \\
Evening type (vs.\ Intermediate) & -0.081 & 0.228 &   442 & -0.354 & 0.723 \\
Social media time          &  0.083 & 0.066 &  1476 &  1.251 & 0.211 \\
Total screen time          & -0.031 & 0.022 &  1835 & -1.410 & 0.159 \\
\bottomrule
\end{tabular}
\vspace{2pt}
\begin{minipage}{0.96\linewidth}
\footnotetext{Sample size: 1918 observations from 461 male participants. \\
Random effects: variance of random intercept (participant) $= 1.558$, $SD = 1.248$; residual variance $= 0.772$, $SD = 0.879$. \\
Model fit: $AIC = 5981.1$; $BIC = 6053.3$; $\mathrm{logLik} = -2977.5$. \\
Intraclass correlation: $ICC = 0.669$. \\
Explained variance: marginal $R^2 = 0.095$; conditional $R^2 = 0.700$.}
\end{minipage}
\end{table}

\begin{table}[htbp]
\centering
\caption{Subgroup analysis: \textbf{Female participants (Mobile)}. 
Linear mixed-effects model predicting UCLA-3 loneliness scores among female participants.
Estimates are unstandardized coefficients.}
\label{tab:subgroup-gender-female-mobile}
\begin{tabular}{lrrrrr}
\toprule
Predictor & Estimate & Std.\ Error & df & $t$ & $p$ \\
\midrule
Intercept                     & 5.274 & 0.361 &   313 & 14.628 & $<0.001$ \\
Wave                          & -0.043 & 0.015 &  1100 & -2.868 & 0.004 \\
Income $>$3000 (vs.\ $\leq$3000) & -0.407 & 0.195 &   267 & -2.089 & 0.038 \\
Age 31--60 (vs.\ 18--30)      & -0.334 & 0.318 &   270 & -1.050 & 0.295 \\
Age 60+ (vs.\ 18--30)         & -0.298 & 0.364 &   273 & -0.820 & 0.413 \\
With partner (vs.\ Single)    & -0.453 & 0.219 &   254 & -2.073 & 0.039 \\
Relationship unknown (vs.\ Single) & 0.142 & 0.242 &   276 &  0.585 & 0.559 \\
Morning type (vs.\ Intermediate)   & -0.284 & 0.189 &   261 & -1.506 & 0.133 \\
Evening type (vs.\ Intermediate)   &  0.094 & 0.275 &   263 &  0.341 & 0.733 \\
Social media time             &  0.133 & 0.058 &  1075 &  2.301 & 0.022 \\
Total screen time             & -0.018 & 0.030 &   954 & -0.607 & 0.544 \\
\bottomrule
\end{tabular}
\vspace{2pt}
\begin{minipage}{0.96\linewidth}
\footnotetext{Sample size: 1360 observations from 271 participants. \\
Random effects: variance of random intercept (participant) $= 1.801$, $SD = 1.342$; residual variance $= 0.849$, $SD = 0.922$. \\
Model fit: $AIC = 4329.2$; $BIC = 4397.0$; $\mathrm{logLik} = -2151.6$. \\
Intraclass correlation: $ICC = 0.680$. \\
Explained variance: marginal $R^2 = 0.080$; conditional $R^2 = 0.705$.}
\end{minipage}
\end{table}

\begin{table}[htbp]
\centering
\caption{Subgroup analysis: \textbf{Female participants (Desktop)}. 
Linear mixed-effects model predicting UCLA-3 loneliness scores among female participants using desktop data.
Estimates are unstandardized coefficients.}
\label{tab:subgroup-gender1-desktop}
\begin{tabular}{lrrrrr}
\toprule
Predictor & Estimate & Std.\ Error & df & $t$ & $p$ \\
\midrule
Intercept                  & 5.180 & 0.263 &   413 & 19.691 & $<0.001$ \\
Wave                       &  0.022 & 0.015 &  1193 &  1.486 & 0.136 \\
Income $>$3000 (vs.\ $\leq$3000) & -0.554 & 0.165 &   375 & -3.347 & $<0.001$ \\
Age 31--60 (vs.\ 18--30)   & -0.444 & 0.238 &   395 & -1.864 & 0.063 \\
Age 60+ (vs.\ 18--30)      & -0.625 & 0.273 &   395 & -2.289 & 0.023 \\
With partner (vs.\ Single) & -0.243 & 0.185 &   365 & -1.316 & 0.189 \\
Relationship unknown (vs.\ Single) &  0.326 & 0.206 &   393 &  1.583 & 0.114 \\
Morning type (vs.\ Intermediate) & -0.380 & 0.157 &   375 & -2.418 & 0.016 \\
Evening type (vs.\ Intermediate) & -0.066 & 0.231 &   398 & -0.285 & 0.776 \\
Social media time          &  0.119 & 0.057 &  1409 &  2.076 & 0.038 \\
Total screen time          & -0.029 & 0.032 &  1227 & -0.909 & 0.364 \\
\bottomrule
\end{tabular}
\vspace{2pt}
\begin{minipage}{0.96\linewidth}
\footnotetext{Sample size: 1500 observations from 390 female participants. \\
Random effects: variance of random intercept (participant) $= 1.763$, $SD = 1.328$; residual variance $= 0.798$, $SD = 0.893$. \\
Model fit: $AIC = 4793.0$; $BIC = 4862.1$; $\mathrm{logLik} = -2383.5$. \\
Intraclass correlation: $ICC = 0.688$. \\
Explained variance: marginal $R^2 = 0.079$; conditional $R^2 = 0.713$.}
\end{minipage}
\end{table}

\begin{table}[htbp]
\centering
\caption{Subgroup analysis: \textbf{Single participants (Mobile)}. 
Linear mixed-effects model predicting UCLA-3 loneliness scores among single participants.
Estimates are unstandardized coefficients.}
\label{tab:subgroup-relationship-single-mobile}
\begin{tabular}{lrrrrr}
\toprule
Predictor & Estimate & Std.\ Error & df & $t$ & $p$ \\
\midrule
Intercept                  & 5.481 & 0.516 &   188 & 10.622 & $<0.001$ \\
Wave                       & -0.051 & 0.017 &   811 & -3.044 & 0.002 \\
Income $>$3000 (vs.\ $\leq$3000) &  0.329 & 0.326 &   166 &  1.009 & 0.314 \\
Female (vs.\ Male)         &  0.049 & 0.238 &   170 &  0.206 & 0.837 \\
Age 31--60 (vs.\ 18--30)   & -0.262 & 0.497 &   169 & -0.527 & 0.599 \\
Age 60+ (vs.\ 18--30)      & -0.001 & 0.556 &   169 & -0.001 & 0.999 \\
Morning type (vs.\ Intermediate)& -0.748 & 0.268 &   166 & -2.793 & 0.006 \\
Evening type (vs.\ Intermediate)& -0.977 & 0.346 &   165 & -2.824 & 0.005 \\
Social media time          &  0.088 & 0.070 &   870 &  1.250 & 0.211 \\
Total screen time          & -0.047 & 0.039 &   876 & -1.211 & 0.226 \\
\bottomrule
\end{tabular}
\vspace{2pt}
\begin{minipage}{0.96\linewidth}
\footnotetext{Sample size: 985 observations from 175 participants. \\
Random effects: variance of random intercept (participant) $= 2.252$, $SD = 1.501$; residual variance $= 0.819$, $SD = 0.905$. \\
Model fit: $AIC = 3118.0$; $BIC = 3176.7$; $\mathrm{logLik} = -1547.0$. \\
Intraclass correlation: $ICC = 0.733$. \\
Explained variance: marginal $R^2 = 0.067$; conditional $R^2 = 0.751$.}
\end{minipage}
\end{table}

\begin{table}[htbp]
\centering
\caption{Subgroup analysis: \textbf{Single participants (Desktop)}. 
Linear mixed-effects model predicting UCLA-3 loneliness scores among single participants using desktop data.
Estimates are unstandardized coefficients.}
\label{tab:subgroup-rel0-desktop}
\begin{tabular}{lrrrrr}
\toprule
Predictor & Estimate & Std.\ Error & df & $t$ & $p$ \\
\midrule
Intercept                  & 5.334 & 0.396 &   276 & 13.466 & $<0.001$ \\
Wave                       &  0.001 & 0.016 &   916 &  0.068 & 0.946 \\
Income $>$3000 (vs.\ $\leq$3000) & -0.394 & 0.283 &   257 & -1.391 & 0.165 \\
Female (vs.\ Male)         & -0.140 & 0.197 &   262 & -0.707 & 0.481 \\
Age 31--60 (vs.\ 18--30)   & -0.247 & 0.377 &   262 & -0.656 & 0.512 \\
Age 60+ (vs.\ 18--30)      & -0.170 & 0.419 &   264 & -0.406 & 0.685 \\
Morning type (vs.\ Intermediate) & -0.678 & 0.220 &   268 & -3.078 & 0.002 \\
Evening type (vs.\ Intermediate) & -0.745 & 0.278 &   264 & -2.678 & 0.008 \\
Social media time          &  0.110 & 0.070 &  1018 &  1.572 & 0.116 \\
Total screen time          & -0.029 & 0.028 &  1128 & -1.006 & 0.315 \\
\bottomrule
\end{tabular}
\vspace{2pt}
\begin{minipage}{0.96\linewidth}
\footnotetext{Sample size: 1144 observations from 268 single participants. \\
Random effects: variance of random intercept (participant) $= 2.227$, $SD = 1.492$; residual variance $= 0.838$, $SD = 0.915$. \\
Model fit: $AIC = 3724.2$; $BIC = 3784.7$; $\mathrm{logLik} = -1850.1$. \\
Intraclass correlation: $ICC = 0.727$. \\
Explained variance: marginal $R^2 = 0.051$; conditional $R^2 = 0.740$.}
\end{minipage}
\end{table}

\begin{table}[htbp]
\centering
\caption{Subgroup analysis: \textbf{With partner (Mobile)}. 
Linear mixed-effects model predicting UCLA-3 loneliness scores among partnered participants.
Estimates are unstandardized coefficients.}
\label{tab:subgroup-relationship-partner-mobile}
\begin{tabular}{lrrrrr}
\toprule
Predictor & Estimate & Std.\ Error & df & $t$ & $p$ \\
\midrule
Intercept                  & 4.106 & 0.413 &   297 &  9.939 & $<0.001$ \\
Wave                       & -0.029 & 0.012 &  1304 & -2.414 & 0.016 \\
Income $>$3000 (vs.\ $\leq$3000) & -0.310 & 0.151 &   277 & -2.057 & 0.041 \\
Female (vs.\ Male)         &  0.178 & 0.150 &   278 &  1.189 & 0.235 \\
Age 31--60 (vs.\ 18--30)   &  0.104 & 0.382 &   277 &  0.273 & 0.785 \\
Age 60+ (vs.\ 18--30)      & -0.031 & 0.409 &   280 & -0.077 & 0.939 \\
Morning type (vs.\ Intermediate)& -0.178 & 0.152 &   276 & -1.174 & 0.241 \\
Evening type (vs.\ Intermediate)&  0.516 & 0.301 &   281 &  1.716 & 0.087 \\
Social media time          &  0.099 & 0.055 &  1148 &  1.780 & 0.075 \\
Total screen time          & -0.013 & 0.026 &  1127 & -0.479 & 0.632 \\
\bottomrule
\end{tabular}
\vspace{2pt}
\begin{minipage}{0.96\linewidth}
\footnotetext{Sample size: 1576 observations from 283 participants. \\
Random effects: variance of random intercept (participant) $= 1.348$, $SD = 1.161$; residual variance $= 0.669$, $SD = 0.818$. \\
Model fit: $AIC = 4582.6$; $BIC = 4646.9$; $\mathrm{logLik} = -2279.3$. \\
Intraclass correlation: $ICC = 0.668$. \\
Explained variance: marginal $R^2 = 0.037$; conditional $R^2 = 0.681$.}
\end{minipage}
\end{table}

\begin{table}[htbp]
\centering
\caption{Subgroup analysis: \textbf{With partner (Desktop)}. 
Linear mixed-effects model predicting UCLA-3 loneliness scores among partnered participants using desktop data.
Estimates are unstandardized coefficients.}
\label{tab:subgroup-rel1-desktop}
\begin{tabular}{lrrrrr}
\toprule
Predictor & Estimate & Std.\ Error & df & $t$ & $p$ \\
\midrule
Intercept                  & 4.683 & 0.282 &   434 & 16.583 & $<0.001$ \\
Wave                       & -0.017 & 0.012 &  1466 & -1.436 & 0.151 \\
Income $>$3000 (vs.\ $\leq$3000) & -0.305 & 0.122 &   400 & -2.503 & 0.013 \\
Female (vs.\ Male)         & 0.158 & 0.121 &   401 &  1.311 & 0.191 \\
Age 31--60 (vs.\ 18--30)   & -0.411 & 0.267 &   412 & -1.540 & 0.124 \\
Age 60+ (vs.\ 18--30)      & -0.689 & 0.288 &   409 & -2.393 & 0.017 \\
Morning type (vs.\ Intermediate) & -0.252 & 0.122 &   400 & -2.057 & 0.040 \\
Evening type (vs.\ Intermediate) & 0.260 & 0.237 &   408 &  1.099 & 0.272 \\
Social media time          & 0.111 & 0.065 &  1718 &  1.714 & 0.087 \\
Total screen time          & -0.028 & 0.025 &  1393 & -1.150 & 0.250 \\
\bottomrule
\end{tabular}
\vspace{2pt}
\begin{minipage}{0.96\linewidth}
\footnotetext{Sample size: 1794 observations from 409 participants. \\
Random effects: variance of random intercept (participant) $= 1.191$, $SD = 1.091$; residual variance $= 0.694$, $SD = 0.833$. \\
Model fit: $AIC = 5321.2$; $BIC = 5387.1$; $\mathrm{logLik} = -2648.6$. \\
Intraclass correlation: $ICC = 0.632$. \\
Explained variance: marginal $R^2 = 0.045$; conditional $R^2 = 0.648$.}
\end{minipage}
\end{table}

\begin{table}[htbp]
\centering
\caption{Subgroup analysis: \textbf{Intermediate type (Mobile)}. 
Linear mixed-effects model predicting UCLA-3 loneliness scores among participants with Intermediate chronotype.
Estimates are unstandardized coefficients.}
\label{tab:subgroup-dmeq-Intermediate-mobile}
\begin{tabular}{lrrrrr}
\toprule
Predictor & Estimate & Std.\ Error & df & $t$ & $p$ \\
\midrule
Intercept                  & 5.527 & 0.362 &   338 & 15.290 & $<0.001$ \\
Wave                       & -0.041 & 0.014 &  1247 & -3.011 & 0.003 \\
Income $>$3000 (vs.\ $\leq$3000) & -0.328 & 0.192 &   295 & -1.709 & 0.089 \\
Female (vs.\ Male)         & -0.041 & 0.173 &   295 & -0.235 & 0.814 \\
Age 31--60 (vs.\ 18--30)   & -0.229 & 0.321 &   307 & -0.715 & 0.476 \\
Age 60+ (vs.\ 18--30)      & -0.061 & 0.375 &   306 & -0.163 & 0.870 \\
With partner (vs.\ Single) & -0.939 & 0.215 &   286 & -4.365 & $<0.001$ \\
Relationship unknown (vs.\ Single) & -0.113 & 0.245 &   311 & -0.462 & 0.645 \\
Social media time          &  0.094 & 0.057 &  1253 &  1.643 & 0.101 \\
Total screen time          & -0.034 & 0.028 &  1333 & -1.186 & 0.236 \\
\bottomrule
\end{tabular}
\vspace{2pt}
\begin{minipage}{0.96\linewidth}
\footnotetext{Sample size: 1538 observations from 305 participants. \\
Random effects: variance of random intercept (participant) $= 1.982$, $SD = 1.408$; residual variance $= 0.784$, $SD = 0.885$. \\
Model fit: $AIC = 4809.2$; $BIC = 4873.3$; $\mathrm{logLik} = -2392.6$. \\
Intraclass correlation: $ICC = 0.717$. \\
Explained variance: marginal $R^2 = 0.106$; conditional $R^2 = 0.747$.}
\end{minipage}
\end{table}

\begin{table}[htbp]
\centering
\caption{Subgroup analysis: \textbf{Intermediate type (Desktop)}. 
Linear mixed-effects model predicting UCLA-3 loneliness scores among Intermediate chronotype participants using desktop data.
Estimates are unstandardized coefficients.}
\label{tab:subgroup-dmeq0-desktop}
\begin{tabular}{lrrrrr}
\toprule
Predictor & Estimate & Std.\ Error & df & $t$ & $p$ \\
\midrule
Intercept                  & 5.228 & 0.284 &   443 & 18.439 & $<0.001$ \\
Wave                       & -0.031 & 0.014 &  1308 & -2.265 & 0.024 \\
Income $>$3000 (vs.\ $\leq$3000) & -0.314 & 0.168 &   407 & -1.876 & 0.061 \\
Female (vs.\ Male)         & 0.069 & 0.150 &   404 &  0.462 & 0.644 \\
Age 31--60 (vs.\ 18--30)   & -0.150 & 0.253 &   423 & -0.592 & 0.554 \\
Age 60+ (vs.\ 18--30)      & -0.238 & 0.302 &   418 & -0.788 & 0.432 \\
With partner (vs.\ Single) & -0.757 & 0.186 &   397 & -4.073 & $<0.001$ \\
Relationship unknown (vs.\ Single) & -0.196 & 0.218 &   425 & -0.899 & 0.369 \\
Social media time          & 0.190 & 0.058 &  1494 &  3.249 & 0.001 \\
Total screen time          & -0.034 & 0.024 &  1511 & -1.374 & 0.170 \\
\bottomrule
\end{tabular}
\vspace{2pt}
\begin{minipage}{0.96\linewidth}
\footnotetext{Sample size: 1655 observations from 412 participants. \\
Random effects: variance of random intercept (participant) $= 1.940$, $SD = 1.393$; residual variance $= 0.826$, $SD = 0.909$. \\
Model fit: $AIC = 5343.0$; $BIC = 5407.9$; $\mathrm{logLik} = -2659.5$. \\
Intraclass correlation: $ICC = 0.701$. \\
Explained variance: marginal $R^2 = 0.082$; conditional $R^2 = 0.726$.}
\end{minipage}
\end{table}

\begin{table}[htbp]
\centering
\caption{Subgroup analysis: \textbf{Morning type (Mobile)}. 
Linear mixed-effects model predicting UCLA-3 loneliness scores among participants with morning type.
Estimates are unstandardized coefficients.}
\label{tab:subgroup-dmeq-morning-mobile}
\begin{tabular}{lrrrrr}
\toprule
Predictor & Estimate & Std.\ Error & df & $t$ & $p$ \\
\midrule
Intercept                  & 4.950 & 0.540 &   233 &  9.167 & $<0.001$ \\
Wave                       & -0.024 & 0.015 &   924 & -1.556 & 0.120 \\
Income $>$3000 (vs.\ $\leq$3000) & -0.244 & 0.169 &   217 & -1.446 & 0.150 \\
Female (vs.\ Male)         &  0.257 & 0.159 &   218 &  1.620 & 0.108 \\
Age 31--60 (vs.\ 18--30)   & -0.642 & 0.495 &   227 & -1.297 & 0.196 \\
Age 60+ (vs.\ 18--30)      & -0.762 & 0.518 &   226 & -1.470 & 0.143 \\
With partner (vs.\ Single) & -0.468 & 0.202 &   210 & -2.320 & 0.021 \\
Relationship unknown (vs.\ Single) & -0.294 & 0.250 &   233 & -1.177 & 0.240 \\
Social media time          &  0.146 & 0.063 &   891 &  2.310 & 0.021 \\
Total screen time          & -0.004 & 0.031 &   667 & -0.129 & 0.898 \\
\bottomrule
\end{tabular}
\vspace{2pt}
\begin{minipage}{0.96\linewidth}
\footnotetext{Sample size: 1132 observations from 222 participants. \\
Random effects: variance of random intercept (participant) $= 1.188$, $SD = 1.090$; residual variance $= 0.721$, $SD = 0.849$. \\
Model fit: $AIC = 3369.3$; $BIC = 3429.7$; $\mathrm{logLik} = -1672.7$. \\
Intraclass correlation: $ICC = 0.622$. \\
Explained variance: marginal $R^2 = 0.066$; conditional $R^2 = 0.647$.}
\end{minipage}
\end{table}

\begin{table}[htbp]
\centering
\caption{Subgroup analysis: \textbf{Morning type (Desktop)}. 
Linear mixed-effects model predicting UCLA-3 loneliness scores among morning type participants using desktop data.
Estimates are unstandardized coefficients.}
\label{tab:subgroup-dmeq1-desktop}
\begin{tabular}{lrrrrr}
\toprule
Predictor & Estimate & Std.\ Error & df & $t$ & $p$ \\
\midrule
Intercept                  & 5.391 & 0.315 &   357 & 17.137 & $<0.001$ \\
Wave                       & 0.027 & 0.014 &  1125 &  1.910 & 0.056 \\
Income $>$3000 (vs.\ $\leq$3000) & -0.435 & 0.136 &   336 & -3.187 & 0.002 \\
Female (vs.\ Male)         & 0.099 & 0.128 &   335 &  0.775 & 0.439 \\
Age 31--60 (vs.\ 18--30)   & -0.994 & 0.289 &   345 & -3.436 & $<0.001$ \\
Age 60+ (vs.\ 18--30)      & -1.319 & 0.306 &   343 & -4.312 & $<0.001$ \\
With partner (vs.\ Single) & -0.346 & 0.160 &   328 & -2.157 & 0.032 \\
Relationship unknown (vs.\ Single) & 0.037 & 0.195 &   362 &  0.191 & 0.849 \\
Social media time          & -0.021 & 0.077 &  1024 & -0.276 & 0.783 \\
Total screen time          & -0.050 & 0.031 &  1315 & -1.619 & 0.107 \\
\bottomrule
\end{tabular}
\vspace{2pt}
\begin{minipage}{0.96\linewidth}
\footnotetext{Sample size: 1409 observations from 347 participants. \\
Random effects: variance of random intercept (participant) $= 1.110$, $SD = 1.053$; residual variance $= 0.692$, $SD = 0.832$. \\
Model fit: $AIC = 4190.0$; $BIC = 4253.0$; $\mathrm{logLik} = -2083.0$. \\
Intraclass correlation: $ICC = 0.616$. \\
Explained variance: marginal $R^2 = 0.099$; conditional $R^2 = 0.654$.}
\end{minipage}
\end{table}

\begin{table}[htbp]
\centering
\caption{Subgroup analysis: \textbf{Evening type (Mobile)}. 
Linear mixed-effects model predicting UCLA-3 loneliness scores among participants with evening type.
Estimates are unstandardized coefficients.}
\label{tab:subgroup-dmeq-evening-mobile}
\begin{tabular}{lrrrrr}
\toprule
Predictor & Estimate & Std.\ Error & df & $t$ & $p$ \\
\midrule
Intercept                  & 4.436 & 0.784 &    64 &  5.660 & $<0.001$ \\
Wave                       & -0.040 & 0.032 &   246 & -1.264 & 0.207 \\
Income $>$3000 (vs.\ $\leq$3000) &  0.179 & 0.469 &    56 &  0.382 & 0.704 \\
Female (vs.\ Male)         &  0.345 & 0.443 &    54 &  0.779 & 0.440 \\
Age 31--60 (vs.\ 18--30)   & -0.246 & 0.703 &    54 & -0.350 & 0.728 \\
Age 60+ (vs.\ 18--30)      & -0.723 & 0.838 &    56 & -0.863 & 0.392 \\
With partner (vs.\ Single) &  0.428 & 0.542 &    53 &  0.791 & 0.433 \\
Relationship unknown (vs.\ Single) &  1.092 & 0.564 &    58 &  1.938 & 0.057 \\
Social media time          &  0.032 & 0.125 &   282 &  0.254 & 0.800 \\
Total screen time          & -0.033 & 0.075 &   290 & -0.437 & 0.662 \\
\bottomrule
\end{tabular}
\vspace{2pt}
\begin{minipage}{0.96\linewidth}
\footnotetext{Sample size: 308 observations from 62 participants. \\
Random effects: variance of random intercept (participant) $= 2.345$, $SD = 1.531$; residual variance $= 0.817$, $SD = 0.904$. \\
Model fit: $AIC = 1001.5$; $BIC = 1046.2$; $\mathrm{logLik} = -488.7$. \\
Intraclass correlation: $ICC = 0.742$. \\
Explained variance: marginal $R^2 = 0.081$; conditional $R^2 = 0.763$.}
\end{minipage}
\end{table}

\begin{table}[htbp]
\centering
\caption{Subgroup analysis: \textbf{Evening type(Desktop)}. 
Linear mixed-effects model predicting UCLA-3 loneliness scores among evening type participants using desktop data.
Estimates are unstandardized coefficients.}
\label{tab:subgroup-dmeq2-desktop}
\begin{tabular}{lrrrrr}
\toprule
Predictor & Estimate & Std.\ Error & df & $t$ & $p$ \\
\midrule
Intercept                  & 4.611 & 0.723 &    98 &  6.376 & $<0.001$ \\
Wave                       & -0.066 & 0.033 &   290 & -1.968 & 0.050 \\
Income $>3000$ (vs.\ $\leq$3000) & -0.491 & 0.366 &    81 & -1.342 & 0.183 \\
Female (vs.\ Male)         & -0.030 & 0.339 &    86 & -0.088 & 0.930 \\
Age 31--60 (vs.\ 18--30)   & -0.034 & 0.648 &    93 & -0.052 & 0.959 \\
Age 60+ (vs.\ 18--30)      & -0.537 & 0.702 &    93 & -0.765 & 0.446 \\
With partner (vs.\ Single) & 0.300 & 0.409 &    83 &  0.732 & 0.466 \\
Relationship unknown (vs.\ Single) & 1.586 & 0.468 &    88 &  3.386 & $0.001$ \\
Social media time          & -0.119 & 0.116 &   330 & -1.022 & 0.307 \\
Total screen time          & 0.048 & 0.063 &   331 &  0.753 & 0.452 \\
\bottomrule
\end{tabular}
\vspace{2pt}
\begin{minipage}{0.96\linewidth}
\footnotetext{Sample size: 354 observations from 92 participants. \\
Random effects: variance of random intercept (participant) $= 1.945$, $SD = 1.395$; residual variance $= 0.945$, $SD = 0.972$. \\
Model fit: $AIC = 1200.8$; $BIC = 1247.2$; $\mathrm{logLik} = -588.4$. \\
Intraclass correlation: $ICC = 0.673$. \\
Explained variance: marginal $R^2 = 0.103$; conditional $R^2 = 0.707$.}
\end{minipage}
\end{table}

\begin{table}[htbp]
\centering
\caption{Subgroup analysis: \textbf{Age group 18--30(Mobile)}. 
Linear mixed-effects model predicting UCLA-3 loneliness scores among young adult participants.
Estimates are unstandardized coefficients.}
\label{tab:subgroup-age0-mobile}
\begin{tabular}{lrrrrr}
\toprule
Predictor & Estimate & Std.\ Error & df & $t$ & $p$ \\
\midrule
Intercept                  & 5.237 & 0.629 &    46 &  8.326 & $<0.001$ \\
Wave                       & -0.041 & 0.043 &   134 & -0.952 & 0.343 \\
Income $>$3000 (vs.\ $\leq$3000) & -0.203 & 0.526 &    30 & -0.386 & 0.702 \\
Female (vs.\ Male)         &  0.307 & 0.523 &    29 &  0.587 & 0.562 \\
With partner (vs.\ Single) & -1.135 & 0.707 &    27 & -1.607 & 0.120 \\
Relationship unknown (vs.\ Single) &  0.408 & 0.636 &    29 &  0.642 & 0.526 \\
Morning type (vs.\ Intermediate) &  0.116 & 0.718 &    29 &  0.162 & 0.873 \\
Evening type (vs.\ Intermediate) & -0.138 & 0.672 &    28 & -0.206 & 0.839 \\
Social media time          & -0.128 & 0.199 &   143 & -0.643 & 0.521 \\
Total screen time          &  0.048 & 0.110 &   160 &  0.437 & 0.663 \\
\bottomrule
\end{tabular}
\vspace{2pt}
\begin{minipage}{0.96\linewidth}
\footnotetext{Sample size: 171 observations from 37 participants. \\
Random effects: variance of random intercept (participant) $= 1.858$, $SD = 1.363$; residual variance $= 0.820$, $SD = 0.905$. \\
Model fit: $AIC = 557.8$; $BIC = 595.5$; $\mathrm{logLik} = -266.9$. \\
Intraclass correlation: $ICC = 0.694$. \\
Explained variance: marginal $R^2 = 0.135$; conditional $R^2 = 0.735$.}
\end{minipage}
\end{table}

\begin{table}[htbp]
\centering
\caption{Subgroup analysis: \textbf{Age group 18--30 (Desktop)}. 
Linear mixed-effects model predicting UCLA-3 loneliness scores among participants aged 18--30 using desktop data.
Estimates are unstandardized coefficients.}
\label{tab:subgroup-age0-desktop}
\begin{tabular}{lrrrrr}
\toprule
Predictor & Estimate & Std.\ Error & df & $t$ & $p$ \\
\midrule
Intercept                  & 5.291 & 0.477 &    78 & 11.082 & $<0.001$ \\
Wave                       & -0.025 & 0.042 &   178 & -0.598 & 0.551 \\
Income $>3000$ (vs.\ $\leq$3000) & -0.527 & 0.367 &    58 & -1.434 & 0.157 \\
Female (vs.\ Male)         & -0.197 & 0.412 &    59 & -0.478 & 0.635 \\
With partner (vs.\ Single) & -0.305 & 0.453 &    54 & -0.673 & 0.504 \\
Relationship unknown (vs.\ Single) & 0.423 & 0.448 &    59 &  0.944 & 0.349 \\
Morning type (vs.\ Intermediate) & 0.359 & 0.405 &    58 &  0.886 & 0.379 \\
Evening type (vs.\ Intermediate) & 0.378 & 0.623 &    68 &  0.607 & 0.546 \\
Social media time          & -0.285 & 0.221 &   214 & -1.289 & 0.199 \\
Total screen time          & -0.009 & 0.094 &   219 & -0.099 & 0.921 \\
\bottomrule
\end{tabular}
\vspace{2pt}
\begin{minipage}{0.96\linewidth}
\footnotetext{Sample size: 233 observations from 67 participants. \\
Random effects: variance of random intercept (participant) $= 1.611$, $SD = 1.269$; residual variance $= 0.968$, $SD = 0.984$. \\
Model fit: $AIC = 797.4$; $BIC = 838.8$; $\mathrm{logLik} = -386.7$. \\
Intraclass correlation: $ICC = 0.625$. \\
Explained variance: marginal $R^2 = 0.081$; conditional $R^2 = 0.655$.}
\end{minipage}
\end{table}

\begin{table}[htbp]
\centering
\caption{Subgroup analysis: \textbf{Age Group 31--60 (Mobile)}. 
Linear mixed-effects model predicting UCLA-3 loneliness scores among midlife adult participants.
Estimates are unstandardized coefficients.}
\label{tab:subgroup-age1-mobile}
\begin{tabular}{lrrrrr}
\toprule
Predictor & Estimate & Std.\ Error & df & $t$ & $p$ \\
\midrule
Intercept                  & 4.833 & 0.166 &   549 & 29.111 & $<0.001$ \\
Wave                       & -0.040 & 0.011 &  1817 & -3.570 & $<0.001$ \\
Income $>$3000 (vs.\ $\leq$3000) & -0.144 & 0.145 &   435 & -0.991 & 0.322 \\
Female (vs.\ Male)         &  0.150 & 0.133 &   435 &  1.127 & 0.260 \\
With partner (vs.\ Single) & -0.561 & 0.170 &   423 & -3.296 & 0.001 \\
Relationship unknown (vs.\ Single) & -0.009 & 0.193 &   458 & -0.048 & 0.962 \\
Morning type (vs.\ Intermediate) & -0.415 & 0.140 &   432 & -2.960 & 0.003 \\
Evening type (vs.\ Intermediate) & -0.078 & 0.229 &   431 & -0.339 & 0.734 \\
Social media time          &  0.123 & 0.045 &  1863 &  2.750 & 0.006 \\
Total screen time          & -0.019 & 0.023 &  1758 & -0.820 & 0.412 \\
\bottomrule
\end{tabular}
\vspace{2pt}
\begin{minipage}{0.96\linewidth}
\footnotetext{Sample size: 2237 observations from 443 participants. \\
Random effects: variance of random intercept (participant) $= 1.714$, $SD = 1.309$; residual variance $= 0.776$, $SD = 0.881$. \\
Model fit: $AIC = 6906.7$; $BIC = 6975.2$; $\mathrm{logLik} = -3441.3$. \\
Intraclass correlation: $ICC = 0.688$. \\
Explained variance: marginal $R^2 = 0.069$; conditional $R^2 = 0.710$.}
\end{minipage}
\end{table}

\begin{table}[htbp]
\centering
\caption{Subgroup analysis: \textbf{Age group 31--60 (Desktop)}. 
Linear mixed-effects model predicting UCLA-3 loneliness scores among participants aged 31--60 using desktop data.
Estimates are unstandardized coefficients.}
\label{tab:subgroup-age1-desktop}
\begin{tabular}{lrrrrr}
\toprule
Predictor & Estimate & Std.\ Error & df & $t$ & $p$ \\
\midrule
Intercept                  & 4.918 & 0.141 &   793 & 34.796 & $<0.001$ \\
Wave                       & -0.027 & 0.011 &  1947 & -2.376 & 0.018 \\
Income $>3000$ (vs.\ $\leq$3000) & -0.274 & 0.130 &   599 & -2.103 & 0.036 \\
Female (vs.\ Male)         & 0.054 & 0.116 &   601 &  0.467 & 0.641 \\
With partner (vs.\ Single) & -0.469 & 0.149 &   587 & -3.141 & 0.002 \\
Relationship unknown (vs.\ Single) & 0.111 & 0.175 &   629 &  0.638 & 0.524 \\
Morning type (vs.\ Intermediate) & -0.427 & 0.123 &   608 & -3.477 & $<0.001$ \\
Evening type (vs.\ Intermediate) & -0.063 & 0.200 &   601 & -0.318 & 0.751 \\
Social media time          & 0.083 & 0.052 &  2243 &  1.599 & 0.110 \\
Total screen time          & -0.035 & 0.020 &  2305 & -1.719 & 0.086 \\
\bottomrule
\end{tabular}
\vspace{2pt}
\begin{minipage}{0.96\linewidth}
\footnotetext{Sample size: 2442 observations from 605 participants. \\
Random effects: variance of random intercept (participant) $= 1.718$, $SD = 1.311$; residual variance $= 0.794$, $SD = 0.891$. \\
Model fit: $AIC = 7725.9$; $BIC = 7795.5$; $\mathrm{logLik} = -3851.0$. \\
Intraclass correlation: $ICC = 0.684$. \\
Explained variance: marginal $R^2 = 0.067$; conditional $R^2 = 0.705$.}
\end{minipage}
\end{table}

\begin{table}[htbp]
\centering
\caption{Subgroup analysis: \textbf{Age Group 60+ (Mobile)}. 
Linear mixed-effects model predicting UCLA-3 loneliness scores among older adult participants.
Estimates are unstandardized coefficients.}
\label{tab:subgroup-age2-mobile}
\begin{tabular}{lrrrrr}
\toprule
Predictor & Estimate & Std.\ Error & df & $t$ & $p$ \\
\midrule
Intercept                  & 5.374 & 0.354 &   124 & 15.196 & $<0.001$ \\
Wave                       & -0.015 & 0.021 &   463 & -0.696 & 0.487 \\
Income $>$3000 (vs.\ $\leq$3000) & -0.800 & 0.319 &   101 & -2.510 & 0.014 \\
Female (vs.\ Male)         &  0.116 & 0.289 &   101 &  0.401 & 0.689 \\
With partner (vs.\ Single) & -0.705 & 0.331 &    98 & -2.130 & 0.036 \\
Relationship unknown (vs.\ Single) & -0.308 & 0.456 &   109 & -0.674 & 0.502 \\
Morning type (vs.\ Intermediate) & -0.635 & 0.302 &   100 & -2.104 & 0.038 \\
Evening type (vs.\ Intermediate) & -0.230 & 0.481 &   103 & -0.479 & 0.633 \\
Social media time          & -0.112 & 0.109 &   478 & -1.025 & 0.306 \\
Total screen time          &  0.014 & 0.051 &   439 &  0.275 & 0.784 \\
\bottomrule
\end{tabular}
\vspace{2pt}
\begin{minipage}{0.96\linewidth}
\footnotetext{Sample size: 570 observations from 109 participants. \\
Random effects: variance of random intercept (participant) $= 1.938$, $SD = 1.392$; residual variance $= 0.695$, $SD = 0.833$. \\
Model fit: $AIC = 1735.3$; $BIC = 1787.4$; $\mathrm{logLik} = -855.6$. \\
Intraclass correlation: $ICC = 0.736$. \\
Explained variance: marginal $R^2 = 0.154$; conditional $R^2 = 0.777$.}
\end{minipage}
\end{table}

\begin{table}[htbp]
\centering
\caption{Subgroup analysis: \textbf{Age group 60+ (Desktop)}. 
Linear mixed-effects model predicting UCLA-3 loneliness scores among participants aged 60+ using desktop data.
Estimates are unstandardized coefficients.}
\label{tab:subgroup-age2-desktop}
\begin{tabular}{lrrrrr}
\toprule
Predictor & Estimate & Std.\ Error & df & $t$ & $p$ \\
\midrule
Intercept                  & 4.728 & 0.281 &   212 & 16.808 & $<0.001$ \\
Wave                       & 0.048 & 0.019 &   591 &  2.513 & 0.012 \\
Income $>3000$ (vs.\ $\leq$3000) & -0.691 & 0.210 &   162 & -3.283 & 0.001 \\
Female (vs.\ Male)         & 0.024 & 0.199 &   164 &  0.123 & 0.903 \\
With partner (vs.\ Single) & -0.590 & 0.227 &   163 & -2.594 & 0.010 \\
Relationship unknown (vs.\ Single) & -0.392 & 0.302 &   182 & -1.298 & 0.196 \\
Morning type (vs.\ Intermediate) & -0.611 & 0.209 &   164 & -2.923 & 0.004 \\
Evening type (vs.\ Intermediate) & -0.157 & 0.309 &   171 & -0.509 & 0.611 \\
Social media time          & 0.207 & 0.082 &   519 &  2.543 & 0.011 \\
Total screen time          & -0.002 & 0.049 &   538 & -0.050 & 0.960 \\
\bottomrule
\end{tabular}
\vspace{2pt}
\begin{minipage}{0.96\linewidth}
\footnotetext{Sample size: 743 observations from 179 participants. \\
Random effects: variance of random intercept (participant) $= 1.351$, $SD = 1.162$; residual variance $= 0.691$, $SD = 0.831$. \\
Model fit: $AIC = 2248.3$; $BIC = 2303.6$; $\mathrm{logLik} = -1112.1$. \\
Intraclass correlation: $ICC = 0.662$. \\
Explained variance: marginal $R^2 = 0.181$; conditional $R^2 = 0.723$.}
\end{minipage}
\end{table}

\clearpage
\section{Social Media Platform Categories}\label{secA5}

\begin{table}[htbp]
\centering
\caption{Classification of social media applications and domains into platform categories.}
\renewcommand{\arraystretch}{1.2}
\begin{tabular}{p{4cm} p{10cm}}
\hline
\textbf{Platform category} & \textbf{Applications / Domains} \\
\hline
Networking-oriented & Facebook, LinkedIn, Nebenan, Twitter \\
Visual-sharing platforms & Instagram, Pinterest, TikTok, YouTube \\
Discussion forums & Finanzfrage, Fotocommunity, Gutefrage, Quora, Reddit, Steam Community \\
Relationship-oriented & Bumble, Finya, Lovoo, Mingle, StayFriends, Tinder \\
Messaging & Discord, Messenger, Skype, Snapchat, Telegram, TextSecure, Threema, VK, WhatsApp \\
\hline
\end{tabular}
\vspace{2pt}
\begin{minipage}{0.96\linewidth}
\footnotetext{The platform categories were used for testing group differences in H3. Only applications or domains used by more than 100 participants were included.}
\label{tab:H3apps}
\end{minipage}
\end{table}

\section{Group differences Results(H2 and H3)}\label{secA6}

\begin{sidewaystable}[htbp]
\centering
\caption{Group differences temporal patterns(H2)}
\footnotesize
\begin{tabular}{lccccccc}
\toprule
\multicolumn{8}{c}{\textbf{Desktop}} \\
\midrule
 & Non-Lonely Mean & Lonely Mean & $n_{\text{non-lonely}}$ & $n_{\text{lonely}}$ & Hedges' $g$ [90\% CI] & Welch's $t$ ($p\_{FDR}$) & Interpretation \\
\midrule
Daily Sessions & 3.11 & 3.72 & 213 & 225 & -0.30 [-0.46, -0.15] & -3.20 (0.006) & \textbf{Significant and non-equivalent*} \\
Session Duration (h) & 6.93 & 4.23 & 213 & 225 & 0.25 [0.11, 0.40] & 2.62 (0.018) & \textbf{Significant and non-equivalent*} \\
Inter-session Interval (h) & 0.58 & 0.71 & 204 & 222 & -0.23 [-0.39, -0.07] & -2.40 (0.023) & Significant and equivalent  \\
Work-Free Difference (h) & 31.73 & 36.31 & 213 & 225 & -0.08 [-0.23, 0.08] & -0.80 (0.424) & Not significant and equivalent \\
\midrule
\multicolumn{8}{c}{\textbf{Mobile}} \\
\midrule

Daily Sessions & 8.10 & 8.61 & 131 & 160 & -0.12 [-0.32, 0.07] & -1.06 (0.579) & Not significant and equivalent \\
Session Duration (h) & 3.15 & 3.28 & 131 & 160 & -0.03 [-0.22, 0.17] & -0.22 (0.994) & Not significant and equivalent \\
Inter-session Interval (h) & 0.80 & 0.83 & 130 & 160 & -0.12 [-0.32, 0.07] & -1.08 (0.579) & Not significant and equivalent \\
Work-Free Difference (h) & 120.14 & 120.20 & 131 & 160 & -0.00 [-0.19, 0.20] & -0.01 (0.994) & Not significant and equivalent \\
\bottomrule
\end{tabular}
\label{tab:H2_Full}
\end{sidewaystable}

\begin{sidewaystable}[htbp]
\centering
\caption{Group differences in type of social media use}
\footnotesize
\begin{tabular}{lccccccc}
\toprule
\multicolumn{8}{c}{\textbf{Desktop}} \\
\midrule
 & Non-Lonely Mean & Lonely Mean & $n_{\text{non-lonely}}$ & $n_{\text{lonely}}$ & Hedges' $g$ [90\% CI] & Welch's $t$ ($p\_{FDR}$) & Interpretation \\
\midrule
Networking-oriented(min) & 6.72 & 12.07 & 213 & 225 & -0.28 [-0.42, -0.14] & -2.99 (0.015) & \textbf{Significant and non-equivalent*} \\
Visual-sharing(min) & 9.61 & 13.02 & 213 & 225 & -0.20 [-0.34, -0.05] & -2.09 (0.094) & Significant and equivalent \\
Relationship-oriented(min) & 0.08 & 0.06 & 213 & 225 & 0.07 [-0.10, 0.21] & 0.69 (0.250) & Not significant and equivalent \\
Messaging(min) & 0.71 & 0.46 & 213 & 225 & 0.14 [-0.01, 0.26] & 1.44 (0.613) & Not significant and equivalent \\
Discussion Forums(min) & 0.43 & 0.48 & 213 & 225 & -0.05 [-0.22, 0.10] & -0.50 (0.618) & Not significant and equivalent \\
\midrule
\multicolumn{8}{c}{\textbf{Mobile}} \\
\midrule

Networking-oriented(min) & 10.93 & 18.35 & 131 & 160 & -0.34 [-0.53, -0.16] & -2.94 (0.004) & \textbf{Significant and non-equivalent*} \\
Visual-sharing(min) & 16.65 & 34.23 & 131 & 160 & -0.47 [-0.62, -0.30] & -4.24 ($<0.001$) & \textbf{Significant and non-equivalent*} \\
Relationship-oriented(min) & 0.95 & 2.11 & 131 & 160 & -0.35 [-0.50, -0.20] & -3.19 (0.003) & \textbf{Significant and non-equivalent*} \\
Messaging(min) & 15.76 & 22.44 & 131 & 160 & -0.36 [-0.51, -0.20] & -3.24 (0.003) & \textbf{Significant and non-equivalent*} \\
Discussion Forums(min) & 0.01 & 0.14 & 131 & 160 & -0.15 [-0.23, -0.03] & -1.38 (0.168) & Not significant and equivalent \\
\bottomrule
\end{tabular}
\label{tab:H3_Full}
\end{sidewaystable}

\clearpage
\section{Sensitivity Analysis Results}\label{secA3}
\begin{table}[htbp]
\centering
\caption{Stability ranking of preprocessing methods across all datasets.}
\label{tab:stability_ranking}
\begin{tabular}{lcccl}
\toprule
\textbf{Method} & \textbf{Average Deviation} & \textbf{Standard Deviation} & \textbf{Ranking} \\
\midrule
Z-score Outlier Removal     & 0.048 & 0.018 & 1 \\
Winsorization $k=2.0$       & 0.056 & 0.030 & 2 \\
Square Root Transform       & 0.060 & 0.023 & 3 \\
Log1p Transform             & 0.066 & 0.025 & 4 \\
Raw Data                    & 0.070 & 0.031 & 5 \\
Winsorization $k=1.5$       & 0.070 & 0.030 & 6 \\
Winsorization $k=1.0$       & 0.088 & 0.035 & 7 \\
\bottomrule
\end{tabular}
\vspace{2pt}
\begin{minipage}{0.96\linewidth}
\footnotetext{Average deviation represents the mean absolute difference between each method's effect sizes and the overall mean effect size per feature, with lower values indicating greater stability. 
Standard deviation shows the variability of stability scores across the four datasets (mobile type, mobile temporal, desktop type, desktop temporal).
Z-score outlier removal consistently demonstrated the most stable effect sizes across all preprocessing approaches, supporting its selection as the primary method for group comparisons.}
\end{minipage}
\end{table}

\begin{figure}[htbp]
    \centering
    \includegraphics[width=0.8\linewidth]{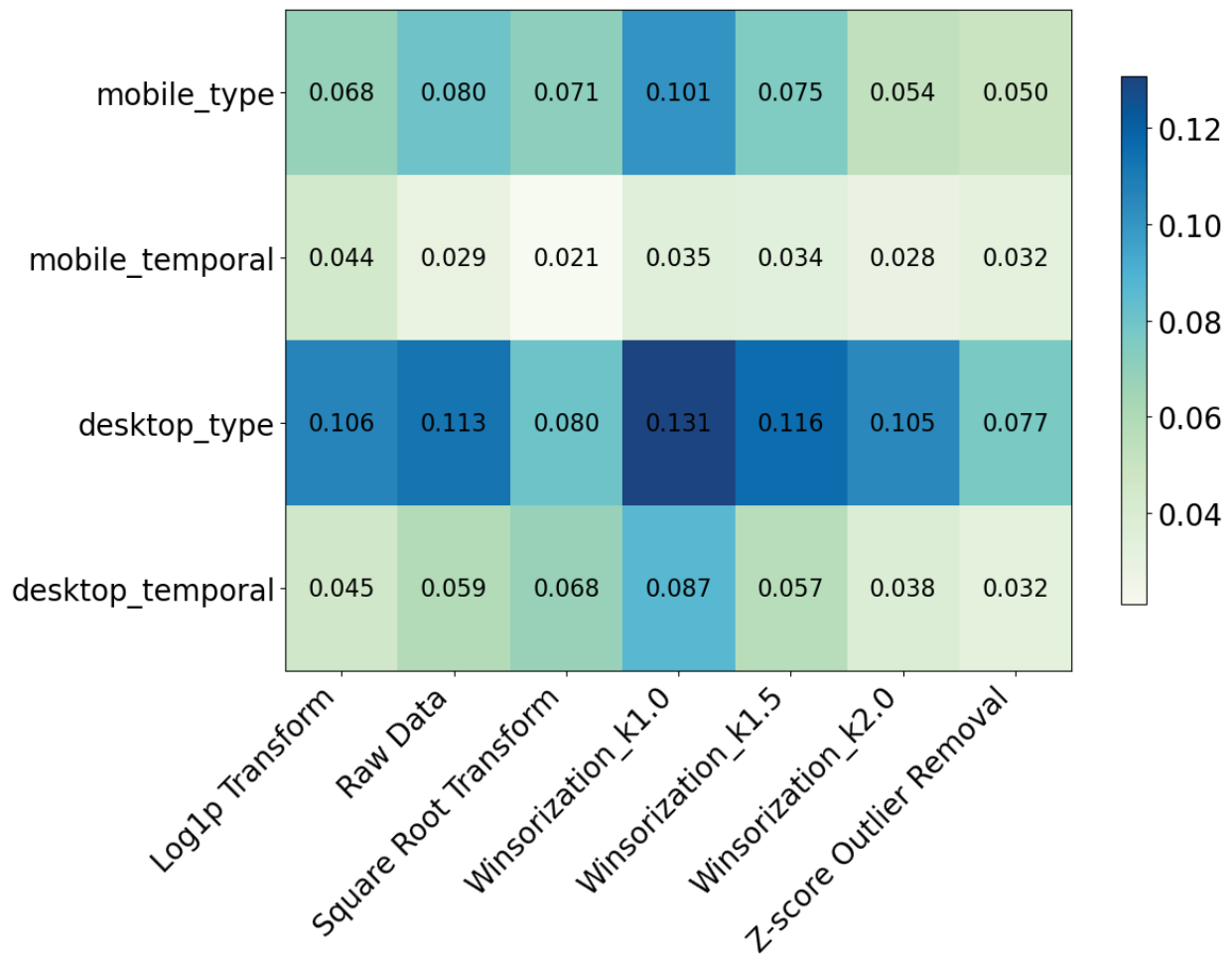}
    \caption{Heat map visualization of preprocessing method stability across datasets and overall method ranking. 
    Method performance is shown by the dataset, with green indicating more stable (lower deviation) methods and red indicating less stable methods. 
    Results support the selection of z-score outlier removal as the primary preprocessing approach for group comparisons.}
    \label{fig:S1}
\end{figure}

\clearpage

\end{appendices}

\end{document}